\documentclass{article}

\usepackage{arxiv}

\usepackage{amsmath}
\usepackage{amssymb}
\usepackage{amsthm}

\usepackage[utf8]{inputenc} 
\usepackage[T1]{fontenc}    
\usepackage{hyperref}       
\usepackage{url}            
\usepackage{booktabs}       
\usepackage{amsfonts}       
\usepackage{nicefrac}       
\usepackage{microtype}      
\usepackage{cleveref}       
\usepackage{lipsum}         
\usepackage{graphicx}
\usepackage{natbib}
\usepackage{chapterbib} 
\usepackage{doi}

\usepackage{amsmath}
\usepackage[american]{babel}
\usepackage{csquotes}
\usepackage{nameref} 

\usepackage{float}

\usepackage[table]{xcolor}
\usepackage{longtable}
\usepackage{makecell}

\newcommand{\manuscripttitle}{Miscalibrated AI improves human calibration: An alignment problem for AI-assisted decision education}
\title{\manuscripttitle}

\renewcommand{\headeright}{Preprint}
\renewcommand{\undertitle}{Preprint}
\renewcommand{\shorttitle}{Miscalibrated AI improves human calibration}

\newif\ifuniqueAffiliation

\ifuniqueAffiliation 
\author{ \href{https://orcid.org/0000-0000-0000-0000}{\includegraphics[scale=0.06]{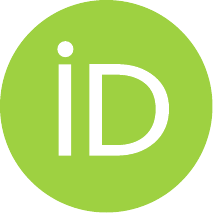}\hspace{1mm}David S.~Hippocampus} \\
	Department of Computer Science\\
	Cranberry-Lemon University\\
	Pittsburgh, PA 15213 \\
	\texttt{hippo@cs.cranberry-lemon.edu} \\
	\And
	\href{https://orcid.org/0000-0000-0000-0000}{\includegraphics[scale=0.06]{orcid.pdf}\hspace{1mm}Elias D.~Striatum} \\
	Department of Electrical Engineering\\
	Mount-Sheikh University\\
	Santa Narimana, Levand \\
	\texttt{stariate@ee.mount-sheikh.edu} \\
}
\else
\usepackage{authblk}

\newbox{\orcid}\sbox{\orcid}{\includegraphics[scale=0.06]{orcid.pdf}} 
\author[1,2]{%
	\href{https://orcid.org/0009-0008-9740-7830}{\usebox{\orcid}\hspace{1mm}Phillip C.~Hegeman\thanks{\href{mailto:pchegeman@gmail.com}\texttt{pchegeman@gmail.com}}}%
}
\author[3]{%
	\href{https://orcid.org/0000-0002-1429-8050}{\usebox{\orcid}\hspace{1mm}Eeshan Hasan}%
}
\author[4]{%
    \href{https://orcid.org/0009-0000-1239-4826}{\usebox{\orcid}\hspace{1mm}Eungik Lee}
}
\author[5]{%
    Andrew Caplin
}
\author[1,2]{%
    \href{https://orcid.org/0000-0002-9019-6574}{\usebox{\orcid}\hspace{1mm}Jennifer S.~Trueblood\thanks{Direct correspondence to \href{mailto:jennifer.s.trueblood@gmail.com}\texttt{jennifer.s.trueblood@gmail.com}}}
}
\affil[1]{Department of Psychological and Brain Sciences, Indiana University Bloomington, IN, USA}
\affil[2]{Cognitive Science Program, Indiana University Bloomington, IN, USA}
\affil[3]{Department of Psychology, The Ohio State University, OH, USA}
\affil[4]{Federal Reserve Bank of New York, NY, USA}
\affil[5]{Department of Economics, New York University, NY, USA}
\fi

\hypersetup{
pdftitle={Miscalibrated AI improves human calibration: An alignment problem for AI-assisted decision education},
pdfsubject={Cognitive science; human--AI interaction},
pdfauthor={Phillip C. Hegeman, Eeshan Hasan, Eungik Lee, Andrew Caplin, Jennifer S. Trueblood},
pdfkeywords={calibration, subjective probability judgment, human--AI interaction, uncertainty communication},
}

\newtheorem{proposition}{Proposition}

\theoremstyle{definition}

\theoremstyle{remark}

\newcommand{\Prb}{\mathbb{P}}
\newcommand{\R}{\mathbb{R}}
\newcommand{\Var}{\mathrm{Var}}

\begin{document}
\maketitle

\begin{abstract}
	Using artificial intelligence (AI) well requires probabilistic thinking, and central to it is \textit{calibration}: expressing certainty that matches one's actual accuracy. Calibration is a decision skill, but a notoriously difficult one to teach. Because AI systems typically generate a probability alongside every prediction, observing these probabilities may help humans learn to provide better calibrated judgments. This is one form of what we call AI-assisted decision education — using AI to teach the judgment skills that decision-making requires — of which learning by observation, as here, is among the simplest forms. In a preregistered experiment, novice participants (N = 300) judged the probability that images of white blood cells were cancerous, before and after exposure to a highly accurate (94\%) convolutional neural network. A single model was shown in three ways, holding accuracy fixed while varying the certainty it expressed: its original probabilities, which were conservative and poorly calibrated for the model's accuracy; a recalibrated version, well calibrated for the model and correspondingly certain; or categorical predictions carrying no probability information. When later tested without the AI, only the original, conservative display improved participants' calibration. It did so by reducing over-extremity, without a condition-specific improvement in discrimination. Notably, the display that taught best was the one miscalibrated for the model. AI-assisted decision education therefore faces an alignment problem: the signal that best describes a model, need not be the signal that best supports the learner.
\end{abstract}


\section{Introduction}
As artificial intelligence (AI) systems are increasingly used to provide decision support in high-stakes domains, it is more critical than ever to ensure that people have the skills to appropriately express, interpret, and integrate uncertainty from many sources \citep{Steyvers.Kumar.2024_ThreeChallengesAIAssisted}. Central among these skills is probability \textit{calibration}. Better-calibrated individuals rely on AI assistance more selectively and appropriately, yielding superior collaborative performance \citep{Ma.et.al.2024_AreYouReally}, and they benefit more from AI assistance compared to their solo performance. This is especially true for those with high calibration but low initial ability, who may have the most to gain from AI assistance \citep{Caplin.et.al.2025_ABCsWhoBenefits}. Calibration also matters upstream of AI deployment: well-calibrated probability judgments from human annotators improve the quality of training data and, in turn, the accuracy and alignment of the AI systems trained on them \citep{Epping.et.al.2026_ImprovingCrowdsourcingAI, Epping.et.al.2025_HarnessingHumanUncertainty}. Improving human calibration - not just accuracy - is therefore a high priority for effective AI development and human-AI collaboration. Deciding when to defer to a model and when to override it---the appropriate reliance that is widely treated as a core competency of AI literacy \citep{Ma.et.al.2024_AreYouReally, Long.Magerko.2020_WhatAILiteracy}---is itself a probabilistic judgment, and so rests on calibration. Probabilistic thinking is in this sense an important target for decision education, and one whose value rises as more decisions are made with AI.

A judge is perfectly calibrated if, across all instances where they assign probability $p$, the outcome occurs $p$ percent of the time \citep{Lichtenstein.et.al.1982_CalibrationProbabilitiesState}. Human judgments are frequently miscalibrated, most commonly exhibiting over-extremity: assigning probabilities more certain (closer to 0 and 1) than their ability to discriminate between true states warrants \citep{Griffin.Brenner.2004_PerspectivesProbabilityJudgment}. Calibration is challenging for AI systems as well, which often require post-processing or fine-tuning to produce reasonably calibrated judgments \citep{Guo.et.al.2017_CalibrationModernNeural}.

In this paper, we test whether AI can help humans learn to provide better calibrated judgments. AI models commonly produce probabilities---an image classifier computes the probability of category membership, a large language model the probability of each possible next token---which carry information about the relationship between an input and the AI's uncertainty, offering insight into underlying difficulty or typicality. If AI uncertainty is made available to a human learner, they might detect these regularities and use them to improve their own calibration. This could circumvent key limitations of prior calibration training, which has been dominated by multifaceted interventions combining expert instruction with personalized feedback \citep{Lichtenstein.Fischhoff.1980_TrainingCalibration, Stone.Opel.2000_TrainingImproveCalibration, Sharp.et.al.1988_PerformanceFeedbackImproves, Kelly.Mandel.2024_EffectCalibrationTraining, Stone.et.al.2023_AutomatedCalibrationTraining}. Such training is difficult to scale: it requires expert pedagogy, often spans multiple sessions \citep{Moore.et.al.2017_ConfidenceCalibrationMultiyear, Gruetzemacher.et.al.2024_CalibrationTrainingImproving}, and needs many trials---often 100 or more---to estimate an individual's miscalibration before feedback can be offered \citep{Kumar.et.al.2019_VerifiedUncertaintyCalibration}. AI probabilities, by contrast, provide a trial-by-trial signal at essentially no marginal cost, and transmit a different kind of information than the trial-by-trial signals that have failed in the past: neither outcome feedback nor proper scoring rules improve calibration \citep{Baranski.Petrusic.1994_CalibrationResolutionConfidence, Martin.Mandel.2025_CalibrationFeedbackPractical}. An AI probability is a model of uncertainty---the degree of certainty expressed by another capable agent, stimulus by stimulus.

We call this possibility \textit{AI-assisted decision education}: using AI not to improve a particular decision, but to teach the judgment skills that decision-making requires — whether through adaptive instruction or, as here, mere exposure to an AI's expressed uncertainty. Research on AI-assisted decision \textit{making} asks how well people decide while the AI is present, and how its displayed confidence should be designed to support decisions \citep{Vodrahalli.et.al.2022_UncalibratedModelsCan, benz2023humanaligned}. Recent work extends this to humans who learn from an AI over repeated interactions, raising the prospect of assistants designed for long-run learning as well as immediate performance \citep{Noti.et.al.2025_AIAssistedDecisionHumanLearning}. Whether such exposure improves people's \textit{unaided} judgments once the AI is withdrawn remains largely untested. Calibration is a demanding test case, given its resistance to conventional training.

There is, however, a fundamental tension in using AI probabilities to teach calibration: the output best calibrated \textit{for the model} may not be the output that best teaches \textit{the learner}. A highly accurate, well-calibrated AI must be quite extreme. Consider an AI classifying images of white blood cells as blasts---immature cells, evidence of cancer when abnormally concentrated---or non-blasts. If it is correct in the vast majority of cases ($>90\%$), well-calibrated output requires high certainty on most trials; at the limit, a perfectly accurate AI's calibrated judgments would be exactly 0\% and 100\%, functionally equivalent to outcome feedback, which is ineffective for teaching calibration \citep{Baranski.Petrusic.1994_CalibrationResolutionConfidence, Martin.Mandel.2025_CalibrationFeedbackPractical}. Worse, for novices already prone to over-extremity, a highly certain AI could reinforce that tendency: \citet{Li.et.al.2025_ConfidenceAlignsUnderstanding} found that AI-assisted decision making persistently shifted self-confidence toward the AI's mean confidence, \textit{worsening} calibration for the initially over-confident majority. Patterns relating stimulus features to certainty may also be easier to learn among more variable predictions spanning the full probability range. In short, model calibration and learner calibration are distinct design targets, and they can conflict.

This tension is an alignment problem at the human--AI interface. An AI's probability output is a statement about the model---how certain it is, given its own accuracy---and simultaneously an intervention on the human who receives it. Alignment concerns this second function: it is a property of the communicated signal relative to the person who uses it, and is distinct from calibration, the signal's fidelity to the model's accuracy. Because the purpose of AI assistance is to serve the human \citep{HadfieldMenell.et.al.2016_CooperativeInverse}, the signal best calibrated to the model need not be the one that best serves the human. For decision making, that signal is tuned to the decision being made now, with the AI present; for decision education, the target is the human's unaided judgment later, once the AI is gone. This is the alignment problem we examine.

This project investigates (a) whether exposure to AI probabilities facilitates learning of probability calibration in human novices, (b) whether any improvements in calibration coincide with improvements in accuracy and judgment informativeness, and (c) whether learning differs by the calibration and extremity of the AI's output. Participants judge the probability that images of white blood cells are blasts (versus non-blasts), with AI predictions provided by a convolutional neural net (CNN) fine-tuned on human-annotated data from this task \citep{Epping.et.al.2025_HarnessingHumanUncertainty}. The design's key feature is that a single underlying AI, highly accurate ($\sim$94\%), was displayed to participants in one of three ways that varied the uncertainty profile while holding accuracy fixed: its \textit{original} probability output, which---because the model was trained on recalibrated human judgments and so expresses a human-like level of certainty rather than one matched to its own accuracy---is conservative and thus poorly calibrated relative to its accuracy; a \textit{recalibrated} version of the same output, well calibrated for the model and correspondingly certain; or \textit{categorical} predictions, a baseline stripped of probability information.

After an initial training phase making subjective probability judgments with trial-by-trial ground-truth feedback, participants practiced with AI predictions displayed after each trial, and were then tested without feedback or AI. Our results show that participants exposed to the original, conservative AI---the display miscalibrated for the model's own ability---learned to respond with less over-extremity, thereby substantially improving their own calibration. AI-assisted decision education is therefore possible. Importantly, the display that taught best was the one poorly calibrated for the model itself. Thus, probability displays optimized for model calibration are not necessarily optimized for human learning.

\section{Results}

\begin{figure*}[htbp!]
    \centering
    \includegraphics[width=0.70\textwidth]{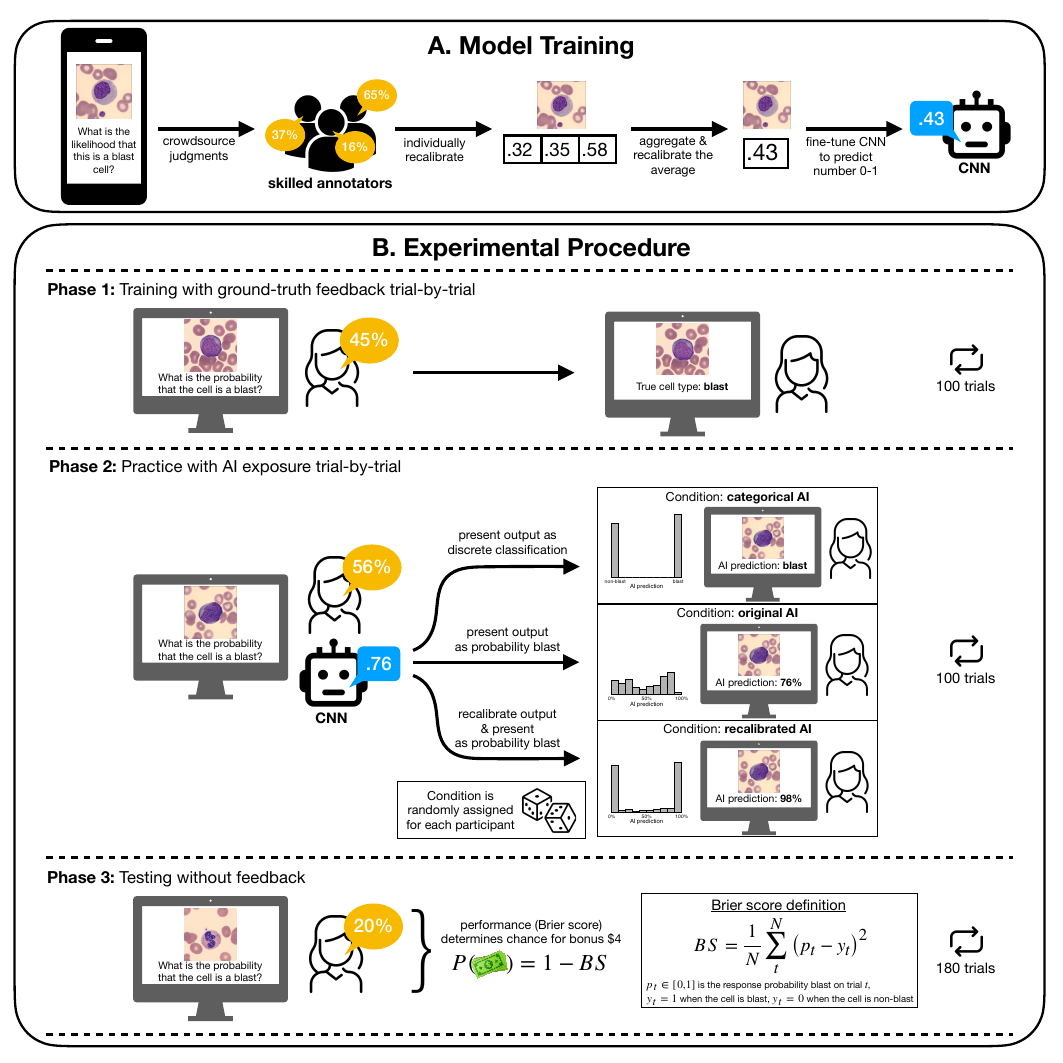}

    \caption{\small \textbf{Study design.} \textbf{A,} AI training procedure from \citet{Epping.et.al.2025_HarnessingHumanUncertainty}: skilled annotators on the DiagnosUs platform provided subjective probabilities that cells were blast, which were aggregated, recalibrated, and used as fine-tuning labels for a convolutional neural network (CNN). \textbf{B,} The three experimental phases. After familiarization with blast and non-blast cells, participants (1) train with trial-by-trial ground-truth feedback, (2) practice with AI predictions shown after every trial, and (3) test without feedback, determining their bonus probability.}

    \label{fig:methods_diagram}
\end{figure*}

In a preregistered experiment, novice participants (N=300) were given an opportunity to study examples of blast (cancer) and non-blast white blood cells before judging the probability that white blood cells were blast across three phases: (1) training with ground truth feedback, (2) practice with AI exposure, and (3) test without feedback or AI (Figure~\ref{fig:methods_diagram}). Participants were given brief instruction in probabilistic judgment and were told that half of the cells in each phase would be blast. Performance in Phase 3 was incentivized by a \$4 bonus with probability determined by their Brier score. AI predictions, coming from a highly accurate (94\%) convolutional neural network (CNN) trained on probability judgments from skilled annotators in \citet{Epping.et.al.2025_HarnessingHumanUncertainty}, varied by condition: original (poorly calibrated, frequently conservative) probabilities, recalibrated (well-calibrated, frequently certain) probabilities, or categorical predictions. Distributions of the output of each AI variant across all 300 stimuli can be seen in Figures \ref{fig:methods_diagram} and S1.

Our analyses compare behavior in pre-AI Phase 1 training (100 trials per participant) to post-AI Phase 3 testing (120 first judgments of non-repeated stimuli). We used Bayesian multilevel linear and logistic regressions with weakly informative or uninformative priors, with by-participant and by-stimulus random effects where appropriate.

\subsection{Performance Metrics}

\begin{figure*}[t]
    \centering
    \includegraphics[width=0.85\textwidth]{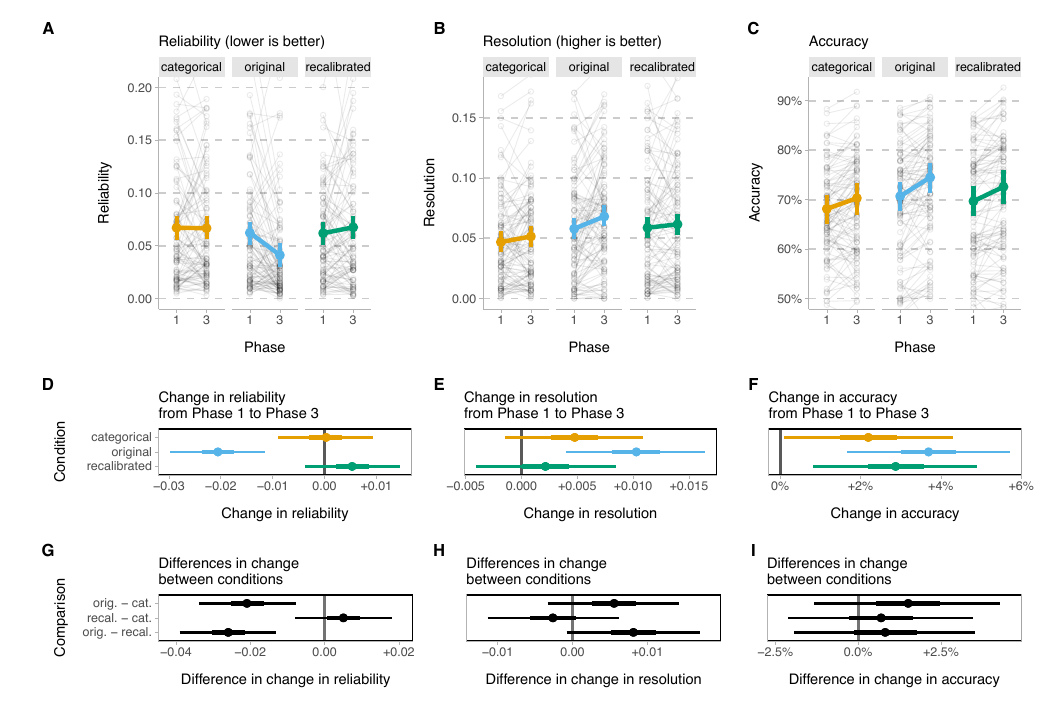} 
    \caption{\textbf{Summary performance metrics.} \textbf{A--C,} Posterior expected reliability (calibration error), resolution (discrimination), and binarized accuracy for each condition and phase, shown as \textit{maximum a posteriori} (MAP) estimate and 95\% HDI; grey points are per-participant MAP estimates. \textbf{D--F,} Expected change in each metric from Phase 1 to 3, with MAP, 50\% HDI (thick) and 95\% HDI (thin). \textbf{G--I,} Expected between-condition difference in that change.}
    \label{fig:reliability_resolution_accuracy}
\end{figure*}

Our central result concerns calibration, so we begin with \textit{reliability}, a component of the Brier score measuring calibration error, where 0 indicates perfect calibration \citep{Murphy.1973_NewVectorPartition}. All three conditions began with mean reliability $\sim$.065 in Phase 1: \textit{maximum a posteriori} estimate (MAP) = .067 (95\% highest density interval, HDI = [.056, .077]) for categorical, .063 ([.051, .073]) for original, and .062 ([.051, .073]) for recalibrated. Reliability was essentially unchanged in Phase 3 for categorical and recalibrated, but improved for original, dropping -.020 ([-.030, -.011]) to .041 ([.030, .053]). For context, the original AI output these participants observed had a reliability of .036, so their Phase 3 calibration approached, but did not reach, that of the display they saw. The original condition improved credibly more than both categorical (difference-in-differences, $\text{DiD}_{O-C}$ = -.021, [-.034, -.008]) and recalibrated ($\text{DiD}_{O-R}$ = -.026, [-.039, -.013]), which did not credibly differ from each other ($\text{DiD}_{R-C}$ = .005, [-.008, .018]). These metrics and their changes are shown in Figure~\ref{fig:reliability_resolution_accuracy}. The same pattern holds for expected calibration error, which decreased only in the original condition (-.046, [-.063, -.030]) and credibly more than in categorical ($\text{DiD}_{O-C}$ = -.048, [-.072, -.025]); see Supplement.

To interpret this calibration gain, we ask whether it was accompanied by improved ability to distinguish blast from non-blast. \textit{Resolution}, another Brier component, is a skill metric ranging from 0 (no discrimination) to .25 (perfect discrimination). All conditions began similarly, though perhaps slightly lower in categorical: .047 (95\% HDI = [.040, .056]) for categorical, .058 ([.049, .066]) for original, and .059 ([.050, .067]) for recalibrated. As with reliability, resolution improved credibly only within the original condition (+.010, [.004, .017]). However, this change did not credibly differ from categorical ($\text{DiD}_{O-C}$ = .006, [-.003, .014]) or recalibrated ($\text{DiD}_{O-R}$ = .008, [-.001, .017]), nor did recalibrated differ from categorical ($\text{DiD}_{R-C}$ = -.002, [-.011, .007]); all intervals include 0.

We also examined classification (``binarized'') accuracy, treating judgments $>.5$ as ``blast'', $<.5$ as ``non-blast'', and exactly $.5$ as a random choice. All conditions began at or just under 70\% in Phase 1: .673 (95\% HDI = [.648, .700]) for categorical, .703 ([.673, .729]) for original, and .692 ([.662, .718]) for recalibrated. All three improved by Phase 3: +.026 ([.005, .044]), +.038 ([.016, .058]), and +.030 ([.010, .049]) respectively, with no evidence of between-condition differences (all 95\% HDIs include 0). The calibration advantage of the original condition therefore cannot be attributed to a condition-specific gain in classification ability.

These results indicate that calibration only improved for the condition exposed to the \textit{worse-calibrated} original AI probabilities, and that this improvement was accompanied by little-to-no relative gain in discrimination. To get a better sense of what might be occurring, we look at the distributions of responses themselves. As can be seen in Figure S1, individuals in all conditions began with a tendency to respond with high degrees of certainty, with almost half of responses being $\leq.1$ or $\geq.9$. For participants exposed to the categorical and recalibrated AI, this tendency persists into Phase 3, essentially unchanged. However, participants exposed to the original AI grow more conservative in Phase 3, responding with probabilities in the most extreme bins in less than 40\% of trials (though, still more frequently than any other bin). This change brings the pooled response distribution of the original AI condition closer in line with the conservatism of the original AI model to which they were exposed.

It is important to note that the response distributions for individual participants vary considerably, with some using near uniform distributions and others consistently selecting responses from a small set of highly certain probabilities. Moreover, the analysis of reliability only summarizes calibration performance without indicating what systematic errors participants are making to produce this level of miscalibration. In order to characterize response and calibration errors while accounting for individual variability, we turn to the linear in log-odds model.

\subsection{Calibration errors}

To identify the precise relationship between probability judgments and relative frequencies---and moreover, \textit{how} this relationship changes to produce improved calibration---we turn to the linear in log-odds model. The linear in log-odds (LLO) model is, in essence, a model-based calibration curve, modeling the "true" probability that a cell is blast as a function of a given subjective probability that it is blast \citep{Turner.et.al.2014_ForecastAggregationRecalibration}. In doing so, it characterizes calibration errors along two dimensions: over-/under-extremity (parameter $\alpha$), and over-/under-prediction (parameter $\beta$). $\alpha < 1$ indicates over-extremity, where subjective probabilities tend to be more certain (closer to 0 and 1) than the corresponding actual relative outcome frequencies. $\alpha > 1$ indicates under-extremity, where subjective probabilities tend to be less certain (closer to .5) than the corresponding relative outcome frequencies. Over- and under-prediction, indicated by $\beta < 0$ and $\beta > 0$, are tendencies for subjective probabilities across the entire [0,1] range to be higher or lower than relative outcome frequencies, respectively. In this case, over-prediction is a sort of systematic bias favoring the probability of blast, and vice versa. Perfect calibration would be indicated by $\alpha$ = 1 and $\beta = 0$.

\begin{figure*}[t]
    \centering
    \includegraphics[width=0.75\textwidth]{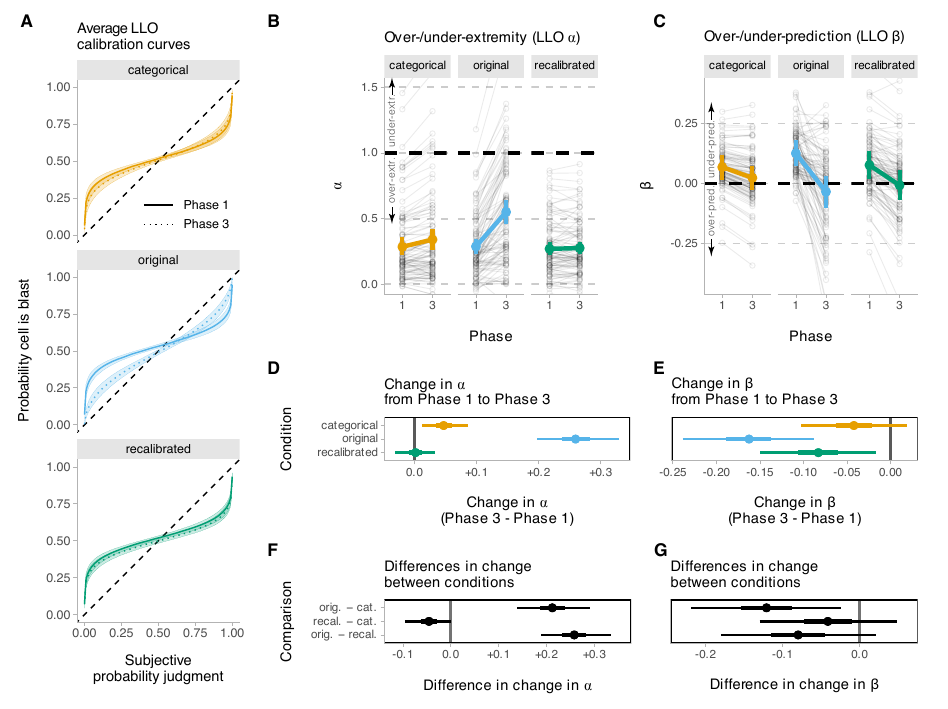} 
    \caption{\textbf{LLO model of calibration errors.} \textbf{A,} LLO-based average calibration curves per condition and phase, showing MAP estimate and 95\% HDI of the expected probability a cell is blast given the subjective probability. \textbf{B,C,} Expected $\alpha$ (over-/under-extremity) and $\beta$ (over-/under-prediction) per condition and phase; grey points are per-participant MAP estimates. \textbf{D,E,} Expected change in $\alpha$ and $\beta$ from Phase 1 to 3, with MAP, 50\% HDI (thick) and 95\% HDI (thin). \textbf{F,G,} Expected between-condition difference in that change.}
    \label{fig:LLO}
\end{figure*}

Estimates of both parameters, and their changes across phases and conditions, are shown in Figure~\ref{fig:LLO}. All conditions begin with considerable over-extremity in Phase 1: mean $\alpha =$ 0.285 (95\% HDI = [0.224, 0.355]) for categorical, 0.291 (95\% HDI = [0.235, 0.344]) for original, 0.275 (95\% HDI = [0.224, 0.326]) for recalibrated. This over-extremity was substantially reduced for the original AI condition, with an increase in $\alpha$ of +0.260 (95\% HDI = [0.195, 0.327]) to an expected value of $\alpha =$ 0.547 (95\% HDI = [0.457, 0.644]) in Phase 3. There is a small but credible increase in average $\alpha$ for the categorical condition (+0.045, 95\% HDI = [0.011, 0.085]), though there is minimal evidence that this change differs from that in the recalibrated condition ($\text{DiD}_{R-C} =$ -0.044, 95\% HDI = [-0.096, 0.002]). By contrast, $\alpha$ increased considerably more in the original condition than in the categorical ($\text{DiD}_{O-C} =$ 0.214, 95\% HDI = [0.137, 0.289]) and the recalibrated conditions ($\text{DiD}_{O-R} =$ 0.256, 95\% HDI = [0.186, 0.333]).

All conditions also exhibited a slight tendency to under-predict the probability that cells were blast in Phase 1: mean $\beta =$ 0.069 (95\% HDI = [0.014, 0.117]) for categorical, 0.126 (95\% HDI = [0.075, 0.181]) for original, and 0.076 (95\% HDI = [0.020, 0.134]) for recalibrated. This under-prediction largely disappeared by Phase 3 for all conditions, with the 95\% HDIs for each condition's Phase 3 mean $\beta$ including 0. There is some evidence of differences between conditions in the change in $\beta$, especially between the original and categorical conditions ($\text{DiD}_{O-C} =$ -0.125, 95\% HDI = [-0.217, -0.022]). However, this difference may partly reflect where the conditions began rather than what they learned. The original AI condition started with a higher mean Phase 1 $\beta$ than the recalibrated condition, by 0.057 (95\% HDI = [-0.013, 0.136]), and so had further to move before reaching $\beta = 0$. Regardless, taken together with the response distributions in Fig.~S1B, which show a high propensity for extreme, certain responses, it is clear that over-extremity is the dominant tendency driving the extent of miscalibration.

\subsection{Blackwell Informativeness}

The reduction in over-extremity we observe for the original AI condition can be achieved in roughly two ways. (1) Individuals apply a \textit{global} adjustment, shifting all judgments to be more conservative so that overall certainty better matches their ability to discriminate. (2) Individuals apportion certainty \textit{differentially across cells}, so that expressed certainty tracks the difficulty of individual stimuli. The two differ in what participants would have learned: the first requires only a uniform shift in the level of certainty, whereas the second requires a stimulus-level mapping from cells to uncertainty. The minimal evidence of relative gains in resolution or accuracy for the original AI condition suggests that (1) plays a considerable role.

Neither resolution nor accuracy can fully adjudicate between the accounts. Binarized accuracy records only which side of .5 a judgment falls on, so a participant could learn which cells are difficult and express that in graded confidence without altering any classification. Resolution does speak to both accounts, but it is an incomplete test: it captures only the spread of the distribution, and only from binned judgments. Two sets of judgments with identical resolution can even support opposite decisions at different intervention thresholds (see Supplement). Distinguishing the accounts thus calls for a criterion defined over the entire distribution rather than a summary of it.

We therefore compare the \textit{Blackwell informativeness} of recalibrated Phase 3 and Phase 1 judgments, a criterion defined over full judgment distributions. For any decision-maker, well-calibrated judgments that are a \textit{mean-preserving spread} of another set are weakly preferable to it \citep{Blackwell.1951_ComparisonExperiments}; if Phase 3 is a mean-preserving spread of Phase 1, Phase 3 judgments are \textit{Blackwell more informative}, supporting universally better decisions.

Recalibration is what makes this comparison diagnostic. A purely global adjustment is undone by the recalibration step: if a participant shrinks every judgment toward .5 by a common transformation, fitting the LLO model recovers that transformation and returns the original recalibrated values. Account (1) therefore predicts no change in Blackwell informativeness by construction, so any gain in spread that survives recalibration must reflect a reallocation of certainty across stimuli, which is account (2).

This yields a clear prediction. Because stimulus-specific uncertainty information was available only in the two probability conditions, learning to apportion certainty more discriminately should produce gains in Blackwell informativeness in the original and recalibrated conditions relative to categorical.

\begin{figure*}[t]
    \centering
    \includegraphics[width=0.70\textwidth]{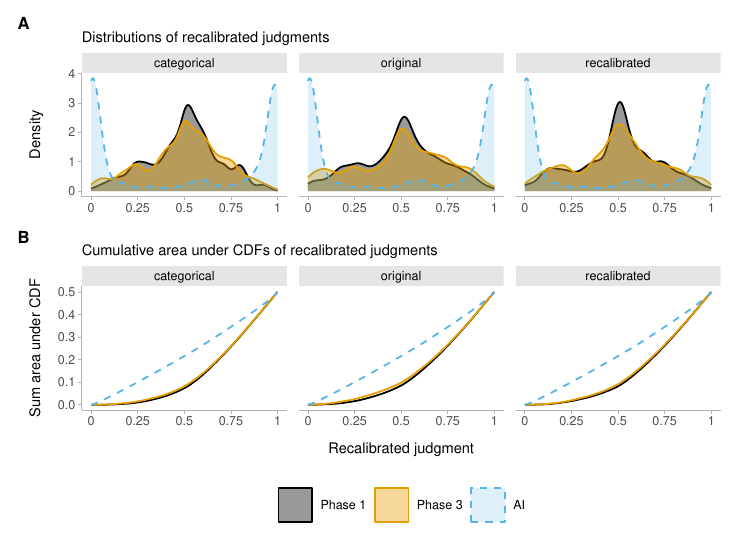} 
    \caption{\textbf{Blackwell informativeness via mean-preserving spread of recalibrated judgments.} \textbf{A,} Pooled distributions of recalibrated judgments per condition in Phases 1 and 3, with recalibrated AI output for reference. Recalibrated probabilities are Bayesian LLO model MAP predictions of the ``true'' probability a cell is blast. \textbf{B,} Cumulative area under the CDF of the distributions in panel A, as a function of recalibrated probability $\tilde{x}$. One distribution is a mean-preserving spread of another, and hence Blackwell more informative, if and only if its cumulative CDF is above the other for all $\tilde{x} \in (0,1)$ (spread) and equal at $\tilde{x} = 1$ (mean-preserving).}
    \label{fig:MPS_recalibrated_distributions}
\end{figure*}

Fig.~\ref{fig:MPS_recalibrated_distributions}A shows the pooled distributions of recalibrated judgments for each phase and condition. All means are approximately equal ($\approx .5$), and the Phase 3 distribution is slightly more dispersed than Phase 1 in all three conditions, with marginally greater density near 0 and 1. The extent of this spread is minimal, especially compared with the AI's recalibrated output, whose density is concentrated near the extremes. That all three conditions shifted slightly in the same direction is consistent with the general improvement in binarized accuracy across conditions, which would produce a small, uniform gain in informativeness irrespective of the AI display.

The mean-preserving spread is evaluated more precisely in Fig.~\ref{fig:MPS_recalibrated_distributions}B, which plots the cumulative area under each empirical CDF. Phase 3 judgments are a mean-preserving spread of Phase 1 judgments if and only if the Phase 1 and Phase 3 lines end with the same y value (i.e., mean-preserving) and the Phase 3 line is above the Phase 1 line for all other points (i.e., spread). Fig. \ref{fig:MPS_recalibrated_distributions}B shows this to be the case, but only barely, and there are no obvious differences between conditions. By contrast, the AI line lies far above both. That gap is itself instructive: participants in the original condition took on much of the AI's level of certainty, without learning to better apportion certainty across cells.

We therefore find no indication of stimulus-level learning: the gain in informativeness is slight and no greater in the two probability conditions, which carried stimulus-specific uncertainty information, than in the categorical condition, which carried none. We note that the criterion is evaluated graphically and the analysis is therefore descriptive rather than inferential. Its value is as converging evidence: the resolution and accuracy analyses likewise show no condition-specific gain in discrimination, and together this is the pattern a global rescaling of certainty would produce. The conceptual and formal basis for this approach is given in the Supplement.

\section{Discussion}

We asked whether exposure to an AI's probability output can teach novices to express better-calibrated uncertainty, and, if so, whether the output that is best calibrated \textit{for the model} is also the output that best teaches \textit{the learner}. The answer to the first question is yes, indicating the feasibility of AI-assisted decision education: simply observing a model's probabilities, with no instruction and no adaptation to the individual, improved a skill that has resisted far more elaborate training. The answer to the second question is no. Only the original, conservative AI---the variant whose probabilities were poorly calibrated relative to the model's own accuracy---improved participants' calibration. The recalibrated variant, which was well calibrated for the model, produced no such improvement, and neither did the categorical baseline.

The improvement was specifically a reduction in \textit{over-extremity}. Participants in every condition began over-extreme, expressing more certainty than their discrimination warranted \citep{Griffin.Brenner.2004_PerspectivesProbabilityJudgment}, and only exposure to the conservative AI lessened this tendency, with the linear-in-log-odds slope $\alpha$ moving toward one. It brought no \textit{relative} gain in discrimination: resolution and binarized accuracy improved similarly across all three conditions, and the gain in Blackwell informativeness was no larger in the two probability conditions than in the categorical baseline, which carried no uncertainty information at all. Thus, participants did not acquire the metacognitive skill of learning \textit{which} cells warrant more certainty. What transferred was a general level of certainty rather than a mapping from cells to certainty, consistent with evidence that people's confidence drifts toward that of an AI they work with \citep{Li.et.al.2025_ConfidenceAlignsUnderstanding}. This also clarifies why the conservative display helped while the certain one did not: novices were already over-extreme, so a highly certain AI---however well calibrated for its own accuracy---offered no corrective example, whereas the conservative AI modeled a degree of uncertainty close to what their performance warranted. The benefit therefore lies in the match between display and learner.

These results connect several literatures. AI probability output is a scalable alternative to conventional calibration training \citep{Lichtenstein.Fischhoff.1980_TrainingCalibration, Stone.et.al.2023_AutomatedCalibrationTraining}, conveying trial-by-trial uncertainty rather than summary feedback---though it may afford only global adjustment without improving the mapping between stimuli and certainty \citep{Keren.1991_CalibrationProbabilityJudgements}. Our findings also complement work on human-aligned AI confidence, which shows that decision-makers integrate AI predictions most effectively when the AI's confidence aligns with their own \citep{benz2023humanaligned, CorveloBenz.GomezRodriguez.2025_HumanalignmentInfluencesUtility}: we provide a learning counterpart, in which an AI whose certainty sits \textit{below} that of an over-extreme learner draws that learner toward better calibration.

Practically, this implies that AI confidence displays may need to be calibrated not only to the model's empirical accuracy but to the learner's current state: a display can be miscalibrated for the AI and still teach a novice well if it models uncertainty appropriate for the learner. For highly accurate systems, whose well-calibrated outputs are necessarily concentrated near zero and one \citep{Guo.et.al.2017_CalibrationModernNeural}, deliberately tempered displays may better support learning, creating a tension between optimizing for the model and for the person. Notably, AI-assisted decisions can be \textit{more} accurate when the AI is made more extreme, eliciting greater trust \citep{Vodrahalli.et.al.2022_UncalibratedModelsCan}---so what is optimal for human--AI collaboration may be the opposite of what is best for human learning.

Several limitations apply. We studied a single perceptual task with novices recruited online, so generalization to other domains and to trained experts remains to be established; the three displays came from a single model at one accuracy level, and whether the pattern holds for less accurate or differently distributed models is unknown. Learning was assessed within the same session, so we cannot speak to durability or transfer. These point to natural extensions: displays might be adapted to each learner, tempering the AI's certainty to their current ability, and tested across domains, expertise levels, and longer retention intervals. Our paradigm was also deliberately passive---participants judged, then observed---which makes it the minimal form of AI-assisted decision education. That minimal form moved calibration but not condition-specific discrimination, suggesting that adaptive instruction, in which an AI responds to a learner's particular errors, may be required for stimulus-level learning.

In short, exposure to miscalibrated AI probability output taught novices to hold beliefs that better matched reality. AI-assisted decision education is therefore feasible, but it inherits an alignment problem of its own: the probability that faithfully represents a model, and the signal that best teaches need not coincide. This does not license covertly misreporting model uncertainty. It implies instead that a model's calibrated probability and a pedagogical assistance signal are distinct design objects: the former should remain available and accurately disclosed, while any transformation of the latter should be explicit, purpose-specific, and governed. Attending to that distinction may be essential to designing AI that helps improve human probabilistic thinking.

\section{Methods}

\subsection{Use of generative AI}

Claude Opus 4.8 and 5 (Anthropic) were used as writing assistants to edit the manuscript. They were not used to generate, analyze, or interpret data, and are not authors. The authors reviewed and verified all text.

\subsection{Experimental Methods}
After familiarization with blast and non-blast images, participants completed three phases of probability judgments: Phase 1 with ground-truth feedback after each judgment, Phase 2 with AI predictions according to assigned condition (categorical, original, or recalibrated), and Phase 3 with no feedback and a performance-based bonus. Comparing Phases 1 and 3 assesses the impact of AI exposure.

\subsubsection{Participants}

Data was collected from $N=300$ participants after exclusions, 100 for each of three between-subjects conditions differing in the kind of AI output they observed in Phase 2 of the experiment:
\begin{enumerate}
    \item \textbf{original} probability output of an overly conservative AI
    \item \textbf{recalibrated} probability output of the same AI, now well-calibrated
    \item \textbf{categorical} predictions from the same AI, serving as a control condition providing binary classifications (i.e., blast, non-blast)
\end{enumerate}

Sample size was based on a previous, similar experiment in which ground-truth feedback and AI predictions were presented together after every training trial rather than in separate phases. That experiment produced credible effects of AI variant with $N=100$ per condition despite divided attention, so we expected the present design to yield effects at least as large and adopted the same sample size.

Participants were U.S. residents recruited through Cloud Research, completing the study online via Amazon Mechanical Turk (MTurk). Participants were paid \$3.00, plus a \$4.00 bonus with chance equal to 1 minus their mean squared error (Brier score) on trials in the final testing phase (Phase 3). All procedures were approved by the Indiana University Human Research Protection Program Institutional Review Board (Protocol \#24488), and all participants provided informed consent before beginning the study. The study was preregistered on the Open Science Framework (\url{https://osf.io/jckpb/overview?view_only=93b1cf3ce2e047e591fca6a129d553d9}).

A total of 327 participants were recruited and completed the experiment. 27 were excluded according to the preregistered criteria (i.e., failing either of two attention check questions or pasting responses for all text response fields), leaving a final sample of $N=300$, with 100 participants in each condition. The attention check questions were "Have you ever used a computer?" ("yes" or "no"), and "I work 14 months in a year." ("true" or "false"). Although we detected 48 participants pasting a text response at least once over the course of the experiment, 44 of them did so on less than four trials, and none pasted on more than 41\% of trials with text fields, so no participants were excluded based on that criterion.

Participants ranged in age from 18 to 77 years ($M = 41.4$, $SD = 11.9$; 57.67\% women, 41.33\% men, 0.33\% other, 0.67\% prefer not to say; 76\% White, 11\% Black or African American, 7\% Asian, 1\% Native American or Pacific Islander, 3\% multiple races selected, 1\% other, 1\% prefer not to say; 9.33\% Hispanic or Latino, 87.33\% non-Hispanic or Latino, 3.33\% other).

The analyses of reliability, resolution, binarized accuracy, and the linear in log-odds model were preregistered. The Blackwell informativeness analysis was not preregistered and is exploratory.

\subsubsection{Materials}

Stimuli consisted of 300 digital images of Wright-stained white blood cells taken from anonymized patient peripheral blood smears at Vanderbilt University Medical Center, as described in \citet{Trueblood.et.al.2018_ImpactSpeedBias}. The images were acquired using a CellaVision DM96 automated digital cell morphology instrument (CellaVision AB, Lund, Sweden).

Ground truth was the unanimous classification of three hematopathology faculty at Vanderbilt University Medical Center, who also rated the typicality of each image on a 1--5 scale; averaged typicality ratings were used to assign images to five difficulty categories from very easy to very hard \citep{Epping.et.al.2026_ImprovingCrowdsourcingAI}. Further detail on classification and typicality rating is given in \citet{Trueblood.et.al.2018_ImpactSpeedBias} and in the Supplement.

For decision training and testing, the present experiment used the subset of 300 images from this larger stimulus bank that were used for machine-learning model training and testing in \citet{Epping.et.al.2025_HarnessingHumanUncertainty}. The remaining 249 images from the original stimulus set were not used as experimental stimuli here because they were used for crowdsourced label recalibration in that prior work; they were, however, used for stimulus familiarization. We selected one of the CNNs trained on the 300-image subset in \citet{Epping.et.al.2025_HarnessingHumanUncertainty} to generate the AI predictions presented to participants.

White blood cell classification provides a useful perceptual learning task because the stimuli are visually complex and vary in difficulty. Blast and non-blast cells differ along multiple morphological dimensions, and the non-blast category is heterogeneous because it includes multiple mature cell types. This complexity makes the task well suited for studying subjective probability judgments and calibration.

\subsubsection{AI model and output variants}

The AI predictions used in this study are from one of the CNNs trained and tested in \citet{Epping.et.al.2025_HarnessingHumanUncertainty}. CNNs tested by Epping et al. varied in how their training data was derived from crowdsourced judgments from skilled annotators. After pre-training on the ImageNet database, the particular CNN we used was fine-tuned to predict the crowdsourced mean subjective probability that a white blood cell is a blast. 

Crowdsourced labels were created by recalibrating individual judgments, averaging by stimulus, and recalibrating the aggregate, which reduced noise and bias and allowed the CNN to learn continuous, human-like uncertainty rather than discrete labels. The selected model was the most accurate of those investigated by \citet{Epping.et.al.2025_HarnessingHumanUncertainty} ($\sim$94\%) while remaining aligned with expert judgments \citep{hasan2024harnessing}, but at a cost to calibration: its probability output is overly conservative. Note that, despite there being equal numbers of blast and non-blast, the AI predicted that 53.7\% of the stimuli were blast (using a decision threshold of .5).

Given the model's high accuracy, a more appropriate calibration would have produced more certain outputs, closer to 0 and 1. We therefore recalibrated the model outputs using a linear-in-log-odds (LLO) model, which models the empirical log-odds of an outcome as a linear function of the probability judgment in log-odds and thus provides a monotonic correction of systematic miscalibration \citep{Turner.et.al.2014_ForecastAggregationRecalibration}. This created an equally accurate but appropriately certain AI variant. The same approach was used to recalibrate crowdsourced judgments in \citet{Epping.et.al.2025_HarnessingHumanUncertainty}, and is described further below.

The LLO transformation was restricted to correct only over- or under-extremity, altering certainty without shifting the decision boundary. The original and recalibrated variants therefore had identical classification accuracy ($\sim$94\%) but differed in apparent certainty. Because the original CNN was primarily under-extreme, recalibration made its outputs more certain, with probabilities $\leq.1$ or $\geq.9$ on over 80\% of stimuli (Fig.~S1), reducing reliability from .036 to .005.

Finally, we created the categorical AI variant by binarizing the original model output probabilities, treating 50\% as a decision threshold such that probabilities $>.5$ become classifications of "blast" and probabilities $<.5$ become classifications of "non-blast".

\subsubsection{Procedure}

The experiment was built using jsPsych \citep{de2023jspsych} and hosted on Pavlovia \citep{Pavlovia}. The primary experimental task was to provide a subjective probability that the white blood cell present in a given image was a \textit{blast} cell (as opposed to a \textit{non-blast} cell). Because participants were expected to be novices, they began by observing 40 examples (20 blast, 20 non-blast), sampled from the set of 249 images in proportion to the overall difficulty distribution and not seen again. Participants were then introduced to the probability judgment task, and the bonus rule was explained qualitatively: greater certainty would improve their chances if correct but sharply worsen them if wrong. Familiarization and instruction details are given in the Supplement.

Participants then completed Phase 1, a training phase of 100 trials (50 blast, 50 non-blast) in randomized order. They were informed that cells were equally likely to be blast or non-blast. On each trial, participants viewed a cell image and entered a probability between 0\% and 100\% that it was a blast, after which the true cell type was displayed alongside their response. Responses and feedback were self-paced. Aggregate performance feedback, in the form of the bonus probability they would have earned, was shown after every 50 trials. Of the 100 trials, 20 came from a common set repeated in every phase (10 blast, 10 non-blast, balanced across difficulty), with the remaining 80 sampled without replacement per participant in proportion to the difficulty distribution. Details are given in the Supplement.

Participants were then introduced to the AI, described as trained to classify cells as blast or non-blast with an accuracy of about 94\%, and told that they would see its predictions instead of the true cell type in Phase 2. Phase 2 comprised 100 trials (50 blast, 50 non-blast) structured as in Phase 1, again with aggregate feedback every 50 trials and the same 20 repeated common stimuli. After each judgment, participants saw the image, their own judgment, and the AI prediction, shown as either a probability rounded to the nearest percent or a categorical classification, depending on condition.

Finally, participants completed Phase 3, a testing phase of 180 trials in randomized order with no feedback of any kind. Of these, 20 were the repeated common set, 120 (60 blast, 60 non-blast) were sampled without replacement in proportion to the difficulty distribution, and 40 were repeats drawn from those 120, used for intra-rater reliability baselines. Stimuli sampled for one phase were removed from the pool for later phases.

After Phase 3, participants completed an exploratory survey on the perceptual attributes underlying their judgments \citep{Tsutsui.et.al.2023_WBCAttWhiteBlood}, their everyday use of AI, and demographics; these responses are not analyzed here. Participants were then told whether they had won the bonus, which 77\% did. Mean completion time was 56.5 minutes (median 48.2, SD 32.1).

\subsection{Analytical Methods}

All analyses share a common approach: comparing Phase 1 to Phase 3 gives the expected change in a variable of interest before and after AI exposure, and comparing those changes across conditions, as in a difference-in-differences analysis, accounts for baseline differences and provides the primary test of whether AI variants produced distinct shifts. Only first judgments of each stimulus were used, giving 100 Phase 1 and 120 Phase 3 trials per participant.

All of the models described below were fit with Markov Chain Monte Carlo (MCMC) in Stan \citep{StanDevelopmentTeam.2025_StanReferenceManual} using the \texttt{CmdStanR} \citep{Gabry.et.al.2025_CmdstanrInterfaceCmdStan} and \texttt{brms} \citep{Burkner.2018_AdvancedBayesianMultilevel} packages in R (v4.4.0). Each model was run using 8 chains with 1,000 tuning iterations and 8,000
post-tuning iterations, for a total of 64,000 post-tuning iterations. Convergence was assessed using $\hat{R}$ ($\hat{R} < 1.01$ indicates convergence) and bulk and tail effective sample size (ESS). Across all models, all parameters and estimands had $\hat{R} < 1.002$, and ESS values were well above conventional thresholds. Weakly informative or uninformative priors were used. Full diagnostics and priors are reported in the Supplement.

\subsubsection{Reliability and Resolution (Calibration Error and Discrimination)}
\label{sec:reliability_resolution_methods}

We assess judgment quality via the Brier score decomposition into uncertainty, reliability, and resolution \citep{Murphy.1973_NewVectorPartition}, which separates how well judgments are calibrated from how well they distinguish blast from non-blast cells. We focus on reliability and resolution, as uncertainty depends only on the overall frequency of blast cells, fixed at 50\% throughout. Details on interpreting reliability and resolution are provided in the Supplement.

\textit{Reliability} is a measure of calibration error: the mean squared difference between subjective probabilities and the actual relative frequencies among trials where that probability is given. Let $x_{j} \in [0,1]$ denote the judgment on stimulus (cell) $j$, with true cell type $y_{j} = 1$ for blast and $y_j = 0$ for non-blast; reliability summarizes the correspondence between $x_j$ and the observed $P_\text{obs}\left(y_j=1 \mid x_j \right)$. Perfect calibration gives a reliability of 0, and the worst possible value of 1 occurs when judgments are completely certain but always wrong.

\textit{Resolution} measures how well judgments distinguish blast from non-blast cells relative to the overall relative frequency: the mean squared difference between the overall frequency $\bar{y}$ and the actual frequencies for different judgments $P_\text{obs}\left(y_j = 1 \mid x_j \right)$. With $\bar{y} = .5$ as in our experiment, resolution ranges from 0, for an individual responding 50\% to every cell and thus providing no information beyond the base rate, to .25, earned when blast and non-blast cells are judged with mutually exclusive probabilities.

For each participant and phase, judgments were divided into five equal-width bins, and within each bin we computed the mean judgment and observed blast frequency. Reliability and resolution were then computed as weighted mean squared differences, with weights proportional to bin counts. Because participants often round to multiples of 5\% or 10\%, bin assignment can depend on whether boundaries are left- or right-inclusive; we therefore averaged metrics across both definitions.

After calculating the reliability and resolution for each participant in Phase 1 and Phase 3 separately, we model each as the outcome in their own Bayesian linear multilevel (mixed-effects) regression. The models include condition and phase as predictors and allow intercepts to vary by participant to account for individual differences in performance that carry across phases. These and all later variable effects are modeled as being normally distributed.

Once a model has been fit, we summarize the posterior by computing the quantity of interest for each posterior sample and reporting its maximum a posteriori (MAP) estimate and 95\% highest density interval (HDI). We did this for the mean predicted metric for each condition at each phase, its change from Phase 1 to Phase 3, and the difference between conditions in that change. This emphasizes estimands that directly address our research questions rather than regression coefficients, which can be difficult to interpret in complex models \citep{Arel-Bundock.2026_ModelMeaningHow}. Parameter estimates are provided in the Supplement.

\subsubsection{Binarized accuracy}
\label{sec:accuracy_methods}

Next, we analyzed binarized accuracy. Probability judgments were coded as “blast” for $x_j>.5$, “non-blast” for $x_j<.5$, and randomly assigned for $x_j=.5$. Judgments were then classified as correct or incorrect based on the true cell type and analyzed using a Bayesian multilevel logistic regression with condition, phase, cell type, and all interactions as predictors. This model tests whether accuracy changed from Phase 1 to Phase 3, whether this change differed by condition, and whether effects varied for blast versus non-blast cells. Correlated random effects were included for participants and stimuli; the full specification is given in the Supplement.

After fitting the accuracy model, we calculated estimands parallel to those for reliability and resolution. Predictions of overall (balanced) accuracy were obtained by averaging the separate blast and non-blast predictions within each posterior sample before computing MAP estimates and HDIs.

\subsubsection{Characterization of calibration errors} \label{sec:LLO_methods}

Reliability captures the overall degree of miscalibration but not its form: whether judgments are more or less certain than warranted (\textit{over-/under-extremity} \citep{Griffin.Brenner.2004_PerspectivesProbabilityJudgment}), or systematically too high or too low (\textit{over-/under-prediction}). To distinguish these, we used the \textit{linear in log-odds} (LLO) model \citep{Turner.et.al.2014_ForecastAggregationRecalibration, Epping.et.al.2026_ImprovingCrowdsourcingAI, Tversky.Fox.1995_WeighingRiskUncertainty, Gonzalez.Wu.1999_ShapeProbabilityWeighting}.

When we analyzed calibration via reliability, we looked at the relationship between judgments $x_j$ and observed relative frequencies $P_\text{obs}(y_{j}=1 \mid x_j)$. The LLO model imposes a particular functional form to the same general relationship. As the name suggests, the LLO model consists of a linear relation between the log-odds of a binary ground truth outcome $P(y_{j}=1)$ and the logit of the subjective probability judgment $x_{j}$:
\begin{equation}\label{LLO_lineareq}
\text{logit}\left(P_\text{LLO}(y_{j}=1 \mid x_j)\right) = \alpha \text{logit}\left(x_{j}\right) + \beta
\end{equation}

The parameters $\alpha, \beta \in (-\infty, \infty)$ capture the systematic calibration errors described above. $\alpha$ captures \textit{over-/under-extremity}: $\alpha > 1$ indicates under-extremity, where participants express less certainty than warranted; $\alpha < 1$ indicates over-extremity, where they express more certainty than warranted; and $\alpha = 1$ indicates no extremity bias. The limiting cases as $\alpha \rightarrow 0$ and below 0 are described in the Supplement.
The parameter $\beta$ captures \textit{over-/under-prediction}, with values of $\beta > 0$ indicating a tendency to under-predict the probability that a cell is blast, and $\beta < 0$ indicating a tendency to over-predict the probability that a cell is blast. An absence of over-/under-prediction bias would be reflected by $\beta = 0$.

The LLO model can be understood as a model-based calibration curve: solving Equation \eqref{LLO_lineareq} for $P_\text{LLO}(y_{j}=1 \mid x_j)$ gives the probability that a cell is blast as a function of the subjective judgment (see Supplement). We also use this model to characterize the (mis)calibration of the original AI output and to construct the recalibrated AI output, as described above.

Because the LLO model has the form of a logistic regression with the logit of the subjective probability as a predictor, we fit it as a Bayesian multilevel logistic regression with true cell type as the outcome and condition, phase, logit judgment, and their interactions as predictors, allowing $\alpha$ and $\beta$ and their change across phases to vary by participant. The full model specification is given in the Supplement.

\subsubsection{Blackwell Informativeness of Recalibrated Subjective Probabilities}
\label{sec:Blackwell_methods}

Resolution summarizes the full distribution of observed relative frequencies by a single number, its spread. To evaluate that distribution as a whole, we employ the Blackwell informativeness framework. \citet{Blackwell.1951_ComparisonExperiments} shows that a decision-maker who maximizes expected utility, whose overall expected utility is therefore a convex function of beliefs, will always prefer well-calibrated judgments that are more dispersed---that is, a \textit{mean-preserving spread}. The conceptual motivation and formal treatment are given in the Supplement.

Our analysis is as follows: (a) recalibrate probability judgments for each individual in each phase, then (b) compare the full distributions to determine whether, and to what degree, Phase 3 recalibrated judgments are a mean-preserving spread of Phase 1 recalibrated judgments in each condition. If this is true, then the judgments at Phase 3 are said to be \textit{Blackwell more informative} than the judgments at Phase 1, supporting better decisions for all payoff functions. If no such relationship holds, then there is no context-free ranking of judgments: some decision payoff structures may favor Phase 1 judgments, while others may favor Phase 3.

We derive recalibrated judgments $\tilde{x}_{ij}$ for each stimulus $j$ judged by participant $i$ from the hierarchical LLO model predictions, $\tilde{x}_{ij} = P_\text{LLO}\left( y_j = 1 \mid i, x_{ij}\right)$, which provide a continuous rather than binned recalibration and reduce the influence of outliers through partial pooling. We then overlay the pooled distributions of MAP recalibrated judgments for Phases 1 and 3 in each condition, together with the recalibrated AI output for reference. As a more precise test, we plot the cumulative area under each distribution's empirical CDF against $\tilde{x}_{ij}$: Phase 3 judgments are a mean-preserving spread of Phase 1 judgments if and only if the two distributions share the same value at $\tilde{x}_j = 1$ and Phase 3 exceeds Phase 1 for all $\tilde{x}_j \in (0,1)$. The full procedure is given in the Supplement.

The conceptual motivation for using recalibrated judgments, and the formal treatment underlying the Blackwell comparison, are provided in the Supplement.

\section{Funding}
A.C. and J.S.T. were funded by the Alfred P. Sloan Foundation.

\bibliographystyle{unsrtnat}
\bibliography{bibliography_bibtex}

\begin{titlepage}
\centering
\vspace*{0.20\textheight}
{\LARGE\bfseries Supplemental Materials: \manuscripttitle\par}
\vspace{2cm}
{\large Phillip C. Hegeman, Eeshan Hasan, Eungik Lee,\par
Andrew Caplin, and Jennifer S. Trueblood\par}
\vspace{1.5cm}
{\normalsize
Department of Psychological and Brain Sciences and Cognitive Science Program, Indiana University Bloomington\par
Department of Psychology, The Ohio State University\par
Federal Reserve Bank of New York\par
Department of Economics, New York University\par}
\end{titlepage}

\setcounter{section}{0}
\setcounter{figure}{0}
\setcounter{table}{0}
\setcounter{equation}{0}
\renewcommand{\thesection}{S\arabic{section}}
\renewcommand{\thefigure}{S\arabic{figure}}
\renewcommand{\thetable}{S\arabic{table}}
\renewcommand{\theequation}{S\arabic{equation}}
\renewcommand{\theHsection}{supp.\arabic{section}}
\renewcommand{\theHfigure}{supp.\arabic{figure}}
\renewcommand{\theHtable}{supp.\arabic{table}}
\renewcommand{\theHequation}{supp.\arabic{equation}}

\makeatletter
\let\articleAddContentsLine\addcontentsline
\renewcommand{\addcontentsline}[3]{%
  \def\contentsFile{#1}\def\mainContentsFile{toc}%
  \ifx\contentsFile\mainContentsFile
    \articleAddContentsLine{stoc}{#2}{#3}%
  \else
    \articleAddContentsLine{#1}{#2}{#3}%
  \fi}
\makeatother
\setcounter{tocdepth}{2}
\makeatletter\@starttoc{stoc}\makeatother
\clearpage


\section{Distributions of Judgments}
\label{sec:histograms_supp}

\begin{figure}[H]
    \centering
    \includegraphics[width=0.65\linewidth]{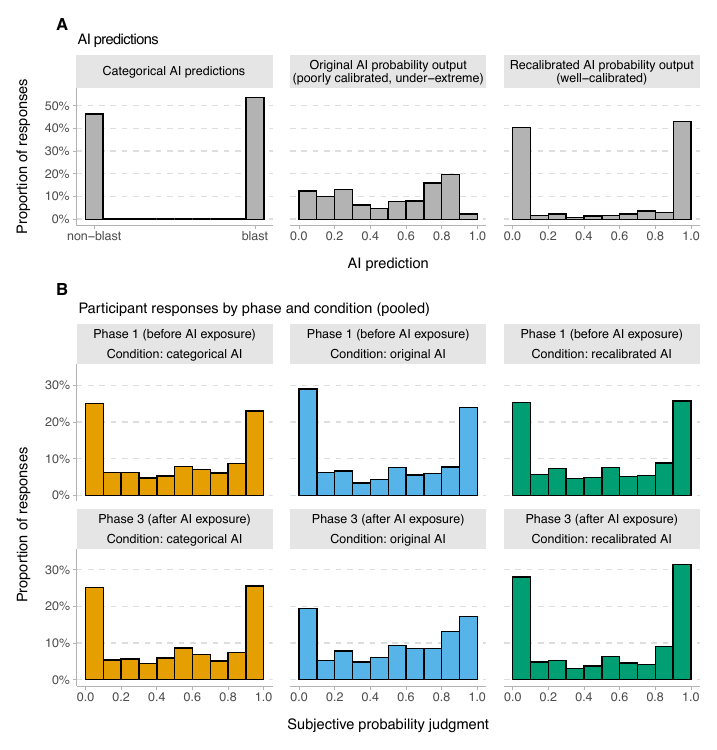}
    \caption{\textbf{Distributions of judgments from AI and human participants.} \textbf{A,} AI output presented in each of the three conditions: categorical (using .5 as the boundary), original AI output, and post-hoc recalibrated AI output. \textbf{B,} Pooled participant responses in Phases 1 and 3 (before and after AI exposure) for each AI exposure condition.}
    \label{fig:histograms_supp}
\end{figure}


\section{Model Specifications and Prior Distributions}
\label{sec:priors}

All models were fit with \texttt{brms} \citep{Burkner.2017_BrmsPackageBayesian}
via the \texttt{cmdstanr} backend \citep{Gabry.et.al.2025_CmdstanrInterfaceCmdStan}.

In the tables below, $\mathcal{N}(\mu,\;\sigma)$ denotes a normal distribution
with mean $\mu$ and standard deviation $\sigma$; $\text{Half-}t(\nu,\;\sigma)$ a half-$t$ distribution (absolute value of a Student's $t$ distribution centered at 0) with degrees of freedom $\nu$ and scale $\sigma$;
$\mathrm{Exp}(\lambda)$ an exponential distribution with rate $\lambda$;
and $\mathrm{LKJ}(\eta)$ the Lewandowski--Kurowicka--Joe prior on correlation matrices.
Condition is dummy-coded with the categorical AI condition as the reference;
phase is dummy-coded with Phase 1 as the reference.

\subsection{Reliability, Resolution, and Expected Calibration Error Models}

The three subject-level metrics were each modeled with a Bayesian linear mixed-effects regression of identical structure.

Let participant $i$ belong to condition $c_i \in \{\text{categorical}, \text{original}, \text{recalibrated}\}$, and let $p \in \{1, 3\}$ denote the phase of the study (pre-AI training or post-AI testing). Define dummy-coded predictors:
\begin{align*}
O_i &=
  \begin{cases}
    1, & \text{for } c_i= \text{original} \\
    0, & \text{otherwise} 
  \end{cases}
  \\
  R_i &=
  \begin{cases}
    1, & \text{for } c_i= \text{recalibrated} \\
    0, & \text{otherwise} 
  \end{cases}
  \\
  P_p &= 
  \begin{cases}
      1, & \text{for phase } p = 3 \text{ (i.e., post-AI testing)}\\
      0, & \text{otherwise (i.e., pre-AI training)}
  \end{cases}
\end{align*}

Then, for each participant $i$, denote the performance metric value (e.g., reliability) in phase $p$ by $y_{ip}$. The model predicting $y_{ip}$ is as follows:
\begin{align*}
  y_{ip}   &\sim \mathcal{N}(\mu_{ip},\;\sigma) \\
  \mu_{ip} &= \beta_0
             + \beta_1 O_i + \beta_2 R_i
             + \beta_3 P_p
             + \beta_4 O_i P_p + \beta_5 R_i P_p
             + u_i \\
  u_i      &\sim \mathcal{N}(0,\;\sigma_\mathrm{participants})
\end{align*}

Priors in all three models were the same except for the intercept $\beta_0$, which represented baseline performance of the reference categorical condition during Phase 1. Weakly informative intercept priors were centered at values consistent with slightly above chance average performance (.25 for reliability, 0 for resolution, .5 for ECE). Priors on baseline condition-differences are narrow, reflecting expectation of negligible pre-existing group differences under random assignment. Priors on all other effects are broad with respect to the scale of the data.
\begin{align*}
    \beta_0 &\sim
        \begin{cases}
            \mathcal{N}(0.2,\;0.3) & \text{for reliability} \\
            \mathcal{N}(0.125,\;0.15) & \text{for resolution} \\
            \mathcal{N}(0.4,\;0.4) & \text{for ECE}
        \end{cases}
    \\
    \beta_1,\;\beta_2 &\sim \mathcal{N}(0,\;0.025)
    \\
    \beta_3,\;\beta_4,\;\beta_5 &\sim \mathcal{N}(0, 0.1)
    \\
    \sigma_\text{participants} &\sim \mathrm{Exp}(2)
    \\
    \sigma &\sim \text{Half-}t(3,\;2.5)
\end{align*}

\subsection{Binarized Accuracy Model}

Binary accuracy was modeled via Bayesian multilevel logistic regression. Keeping the same notation from the previous section, further denote $\mathrm{correct}_{ipt} \in \{0,1\}$ as the correctness of trial $t$ involving participant $i$ judging stimulus $s_t$ of cell type $k \in \{\text{blast}, \text{non-blast}\}$, in phase $p$. The model predicting $\mathrm{correct}_{ipt}$ is as follows:
\begin{align*}
  \mathrm{correct}_{ipt}  &\sim \mathrm{Bernoulli}(\pi_{ipt}) \\
  \mathrm{logit}(\pi_{ipt})
    &= \beta_0
       + \beta_1 O_i + \beta_2 R_i
       + \beta_3 P_p
       + \beta_4 O_i P_p + \beta_5 R_i P_p
       + u_{i \mid c_i}
       + v_{s_t \mid k}
\end{align*}
Participant-level random effects are estimated \emph{separately within each condition}
and include correlated intercepts and Phase~3 slopes:
\[
  \bigl(u_0,\; u_\mathrm{ph3}\bigr)_{i;\,c}
  \;\sim\;
  \mathrm{Normal}\!\bigl(\mathbf{0},\;\boldsymbol{\Sigma}^{(\mathrm{part})}_{c}\bigr),
  \quad c \in \{\mathrm{cat.},\; \mathrm{orig.},\; \mathrm{recal.}\}.
\]
Stimulus-level random effects are estimated \emph{separately within each cell type}
and include correlated intercepts and slopes for condition, phase, and their interaction:
\[
  \mathbf{v}_{s;\,k}
  \;\sim\;
  \mathrm{Normal}\!\bigl(\mathbf{0},\;\boldsymbol{\Sigma}^{(\mathrm{stim})}_{k}\bigr),
  \quad k \in \{\mathrm{blast},\; \mathrm{non\text{-}blast}\}.
\]
Priors are given in Supplementary Table~\ref{tab:priors_correctness}.

\begin{table}[H]
\centering
\caption{Prior distributions for the classification accuracy model.
  All random-effect SD parameters share a single prior regardless of group or slope.}
\label{tab:priors_correctness}
\begin{tabular}{lc}
\toprule
\textbf{Parameter} & \textbf{Prior} \\
\midrule
\multicolumn{2}{l}{\textit{Fixed effects}} \\[2pt]
\quad Intercept                          & $\mathcal{N}(0.4,\ 0.5)$ \\[3pt]
\quad Cond.\ orig.\ (vs cat.)            & $\mathcal{N}(0,\ 0.3)$    \\[3pt]
\quad Cond.\ recal.\ (vs cat.)           & $\mathcal{N}(0,\ 0.3)$    \\[3pt]
\quad Phase 3 (vs 1)                     & $\mathcal{N}(0,\ 1)$    \\[3pt]
\quad Cond.\ orig.\ $\times$ Phase 3     & $\mathcal{N}(0,\ 1)$    \\[3pt]
\quad Cond.\ recal.\ $\times$ Phase 3    & $\mathcal{N}(0,\ 1)$    \\[3pt]
\midrule
\multicolumn{2}{l}{\textit{Random effect SDs}} \\[2pt]
\quad Group: participant (in categorical AI cond.)  & $\mathrm{Exp}(2)$ \\[3pt]
\quad Group: participant (in original AI cond.)     & $\mathrm{Exp}(2)$ \\[3pt]
\quad Group: participant (in recalibrated AI cond.) & $\mathrm{Exp}(2)$ \\[3pt]
\quad Group: stimulus (non-blast)                   & $\mathrm{Exp}(2)$ \\[3pt]
\quad Group: stimulus (blast)                       & $\mathrm{Exp}(2)$ \\[3pt]
\midrule
\multicolumn{2}{l}{\textit{Correlation matrices}} \\[2pt]
\quad Group: participant (all conditions) & $\mathrm{LKJ}(1)$ \\[3pt]
\quad Group: stimulus (all cell types)   & $\mathrm{LKJ}(1)$ \\
\bottomrule
\end{tabular}
\end{table}

\noindent\small\textit{Note.}
Each \textit{Group: participant} SD prior covers two within-condition parameters
(intercept SD and Phase~3 slope SD).
Each \textit{Group: stimulus} SD prior covers six within-cell-type parameters:
the intercept, both condition-dummy slopes, the Phase~3 slope, and both
condition $\times$ Phase~3 interaction slopes.
$\mathrm{LKJ}(1)$ is the uniform distribution over all positive-definite
correlation matrices of the appropriate dimension (\texttt{brms} default).

\subsection{LLO Calibration Model}

The LLO model uses Bayesian multilevel logistic regression with binary cell type as
the outcome. Let $y_t \in \{0,1\}$ denote ground-truth cell type for trial $t$,
and $\ell_t = \mathrm{logit}(x_t)$ the logit-transformed subjective probability judgment.
The linear predictor decomposes into terms determining $\beta$
(over-/under-prediction) and terms determining $\alpha$ (over-/under-extremity),
each modulated by condition and phase:
\begin{align*}
  y_t &\sim \mathrm{Bernoulli}(\pi_t) \\[6pt]
  \mathrm{logit}(\pi_t)
    &= \underbrace{
         \beta_0 + \beta_1 C_{\text{orig.},i} + \beta_2 C_{\text{recal.},i}
         + \beta_3 P_{3,i} + \beta_4 (C_{\text{orig.},i} P_{3,i}) + \beta_5 (C_{\text{recal.},i} P_{3,i})
       }_{\text{determines }\beta} \\[10pt]
    &\quad +
       \underbrace{
         \bigl[
           \beta_6 + \beta_7 C_{\text{orig.},i} + \beta_8 C_{\text{recal.},i}
           + \beta_9 P_{3,i} + \beta_{10} (C_{\text{orig.},i} P_{3,i}) + \beta_{11} (C_{\text{recal.},i} P_{3,i})
         \bigr]\,\ell_t
       }_{\text{determines }\alpha}
       \;+\; u^{(\mathrm{part})}_{i;\,c_i}
\end{align*}
Participant-level random effects are estimated \emph{separately within each condition}
and include correlated intercepts and slopes for phase, logit judgment,
and their interaction:
\[
  \bigl(u_0,\; u_\mathrm{ph3},\; u_\ell,\; u_{\mathrm{ph3} \times \ell}\bigr)_{i;\,c}
  \;\sim\;
  \mathrm{Normal}\!\bigl(\mathbf{0},\;\boldsymbol{\Sigma}^{(\mathrm{part})}_{c}\bigr),
  \quad c \in \{\mathrm{cat.},\; \mathrm{orig.},\; \mathrm{recal.}\}.
\]
Priors are given in Supplementary Table~\ref{tab:priors_LLO}.

\begin{table}[H]
\centering
\caption{Prior distributions for the LLO calibration model.
  Fixed effects are organized by their role in determining $\beta$
  (over-/under-prediction; non-$\ell_t$ terms) or $\alpha$
  (over-/under-extremity; $\ell_t$ terms).}
\label{tab:priors_LLO}
\begin{tabular}{lc}
\toprule
\textbf{Parameter} & \textbf{Prior} \\
\midrule
\multicolumn{2}{l}{\textit{Fixed effects---$\beta$ (over-/under-prediction) parameters}} \\[2pt]
\quad Intercept                          & $\mathcal{N}(0,\ 1)$ \\[3pt]
\quad Cond.\ orig.\ (vs cat.)            & $\mathcal{N}(0,\ 0.3)$ \\[3pt]
\quad Cond.\ recal.\ (vs cat.)           & $\mathcal{N}(0,\ 0.3)$ \\[3pt]
\quad Phase 3 (vs 1)                     & $\mathcal{N}(0,\ 1)$ \\[3pt]
\quad Cond.\ orig.\ $\times$ Phase 3     & $\mathcal{N}(0,\ 1)$ \\[3pt]
\quad Cond.\ recal.\ $\times$ Phase 3    & $\mathcal{N}(0,\ 1)$ \\[3pt]
\midrule
\multicolumn{2}{l}{\textit{Fixed effects---$\alpha$ (over-/under-extremity) parameters}} \\[2pt]
\quad Logit judgment
  & $\mathcal{N}(0.8,\ 1)$ \\[3pt]
\quad Cond.\ orig.\ $\times$ Logit judgment
  & $\mathcal{N}(0,\ 0.3)$ \\[3pt]
\quad Cond.\ recal.\ $\times$ Logit judgment
  & $\mathcal{N}(0,\ 0.3)$ \\[3pt]
\quad Phase 3 $\times$ Logit judgment
  & $\mathcal{N}(0,\ 1)$ \\[3pt]
\quad Cond.\ orig.\ $\times$ Phase 3 $\times$ Logit judgment
  & $\mathcal{N}(0,\ 1)$ \\[3pt]
\quad Cond.\ recal.\ $\times$ Phase 3 $\times$ Logit judgment
  & $\mathcal{N}(0,\ 1)$ \\[3pt]
\midrule
\multicolumn{2}{l}{\textit{Random effect SDs}} \\[2pt]
\quad Group: participant (in categorical AI cond.)  & $\mathrm{Exp}(2)$ \\[3pt]
\quad Group: participant (in original AI cond.)     & $\mathrm{Exp}(2)$ \\[3pt]
\quad Group: participant (in recalibrated AI cond.) & $\mathrm{Exp}(2)$ \\[3pt]
\midrule
\multicolumn{2}{l}{\textit{Correlation matrices}} \\[2pt]
\quad Group: participant (all conditions) & $\mathrm{LKJ}(1)$ \\
\bottomrule
\end{tabular}
\end{table}

\noindent\small\textit{Note.}
The logit judgment parameter ($\beta_6$ above) corresponds to baseline $\alpha$
in the categorical AI condition at Phase~1 and is centered at $0.80$ rather than
$1.00$ (perfect calibration), reflecting the well-established empirical tendency
toward over-extremity in probability judgments.
Each \textit{Group: participant} SD prior covers all four within-condition slopes
(intercept, Phase~3, logit judgment, and Phase~3 $\times$ logit judgment).
$\mathrm{LKJ}(1)$ is the uniform distribution over all positive-definite correlation
matrices of the appropriate dimension (\texttt{brms} default).

\section{Performance metric models}

\subsection{Reliability}

\begin{table}[H]
\centering
\begin{tabular}{llrlrrr}
\toprule
Condition & Phase & MAP & 95\% HDI & $\hat{R}$ & Bulk-ESS & Tail-ESS\\
\midrule
\cellcolor{gray!10}{categorical} & \cellcolor{gray!10}{1} & \cellcolor{gray!10}{0.067} & \cellcolor{gray!10}{{}[.056, .077]} & \cellcolor{gray!10}{1.000} & \cellcolor{gray!10}{28347} & \cellcolor{gray!10}{40393}\\
categorical & 3 & 0.067 & {}[.056, .078] & 1.000 & 33447 & 44745\\
\cellcolor{gray!10}{original} & \cellcolor{gray!10}{1} & \cellcolor{gray!10}{0.063} & \cellcolor{gray!10}{{}[.051, .073]} & \cellcolor{gray!10}{1.000} & \cellcolor{gray!10}{35709} & \cellcolor{gray!10}{47090}\\
original & 3 & 0.041 & {}[.030, .053] & 1.000 & 36571 & 48995\\
\cellcolor{gray!10}{recalibrated} & \cellcolor{gray!10}{1} & \cellcolor{gray!10}{0.062} & \cellcolor{gray!10}{{}[.051, .073]} & \cellcolor{gray!10}{1.000} & \cellcolor{gray!10}{36457} & \cellcolor{gray!10}{47322}\\
\addlinespace
recalibrated & 3 & 0.068 & {}[.056, .078] & 1.000 & 37223 & 49287\\
\bottomrule
\end{tabular}
    \caption{Posterior expected value of a participants' \textbf{reliability} (i.e., Brier score decomposition measure of calibration error) for each condition in each phase.}
    \label{tab:reliability_prediction}
\end{table}

\begin{table}[H]
\centering
\begin{tabular}{lrlrrr}
\toprule
Condition & MAP & 95\% HDI & $\hat{R}$ & Bulk-ESS & Tail-ESS\\
\midrule
\cellcolor{gray!10}{categorical} & \cellcolor{gray!10}{0.000} & \cellcolor{gray!10}{{}[-.009, .009]} & \cellcolor{gray!10}{1.000} & \cellcolor{gray!10}{79992} & \cellcolor{gray!10}{52492}\\
original & -0.020 & {}[-.030, -.011] & 1.000 & 81302 & 55565\\
\cellcolor{gray!10}{recalibrated} & \cellcolor{gray!10}{0.005} & \cellcolor{gray!10}{{}[-.004, .014]} & \cellcolor{gray!10}{1.000} & \cellcolor{gray!10}{81817} & \cellcolor{gray!10}{54005}\\
\bottomrule
\end{tabular}
    \caption{Posterior expected \textbf{change in reliability} for each condition from Phase 1 to 3.}
    \label{tab:reliability_change}
\end{table}

\begin{table}[H]
\centering
\begin{tabular}{lrlrrr}
\toprule
Comparison & MAP & 95\% HDI & $\hat{R}$ & Bulk-ESS & Tail-ESS\\
\midrule
\cellcolor{gray!10}{original - categorical} & \cellcolor{gray!10}{-0.021} & \cellcolor{gray!10}{{}[-.034, -.008]} & \cellcolor{gray!10}{1.000} & \cellcolor{gray!10}{86288} & \cellcolor{gray!10}{56141}\\
recalibrated - categorical & 0.005 & {}[-.008, .018] & 1.000 & 86938 & 55048\\
\cellcolor{gray!10}{original - recalibrated} & \cellcolor{gray!10}{-0.026} & \cellcolor{gray!10}{{}[-.039, -.013]} & \cellcolor{gray!10}{1.000} & \cellcolor{gray!10}{100666} & \cellcolor{gray!10}{42918}\\
\bottomrule
\end{tabular}
    \caption{Posterior expected \textbf{difference between conditions in the change in reliability} from Phase 1 to Phase 3.}
    \label{tab:reliability_difinchange}
\end{table}

\begin{table}[H]
\centering
\begin{tabular}{lrlrrr}
\toprule
Parameter & MAP & 95\% HDI & $\hat{R}$ & Bulk-ESS & Tail-ESS\\
\midrule
\addlinespace[0.3em]
\multicolumn{6}{l}{\textbf{Fixed Effects $\beta$s}}\\
\hspace{1em}\cellcolor{gray!10}{Intercept} & \cellcolor{gray!10}{0.067} & \cellcolor{gray!10}{{}[.056, .077]} & \cellcolor{gray!10}{1.000} & \cellcolor{gray!10}{28347} & \cellcolor{gray!10}{40393}\\
\hspace{1em}Cond. orig. (vs cat.) & -0.006 & {}[-.019, .010] & 1.000 & 30713 & 39691\\
\hspace{1em}\cellcolor{gray!10}{Cond. recal. (vs cat.)} & \cellcolor{gray!10}{-0.004} & \cellcolor{gray!10}{{}[-.019, .011]} & \cellcolor{gray!10}{1.000} & \cellcolor{gray!10}{29971} & \cellcolor{gray!10}{42149}\\
\hspace{1em}Phase 3 (vs 1) & 0.000 & {}[-.009, .009] & 1.000 & 79992 & 52492\\
\hspace{1em}\cellcolor{gray!10}{Cond. orig. (vs cat.) × Phase 3 (vs 1)} & \cellcolor{gray!10}{-0.021} & \cellcolor{gray!10}{{}[-.034, -.008]} & \cellcolor{gray!10}{1.000} & \cellcolor{gray!10}{86288} & \cellcolor{gray!10}{56141}\\
\hspace{1em}Cond. recal. (vs cat.) × Phase 3 (vs 1) & 0.005 & {}[-.008, .018] & 1.000 & 86938 & 55048\\
\addlinespace[0.3em]
\multicolumn{6}{l}{\textbf{Residual SD}}\\
\hspace{1em}\cellcolor{gray!10}{Sigma} & \cellcolor{gray!10}{0.033} & \cellcolor{gray!10}{{}[.030, .036]} & \cellcolor{gray!10}{1.000} & \cellcolor{gray!10}{31246} & \cellcolor{gray!10}{47067}\\
\addlinespace[0.3em]
\multicolumn{6}{l}{\textbf{Random Effect SDs}}\\
\addlinespace[0.3em]
\multicolumn{6}{l}{\textbf{\quad Group: participant}}\\
\hspace{1em}\hspace{1em}Intercept & 0.046 & {}[.042, .051] & 1.000 & 21449 & 38219\\
\bottomrule
\end{tabular}
    \caption{Population-level parameters from linear mixed-effects model of participant reliability.}
    \label{tab:reliability_parameters}
\end{table}

\subsection{Resolution}

\begin{table}[H]
\centering
\begin{tabular}{llrlrrr}
\toprule
Condition & Phase & MAP & 95\% HDI & $\hat{R}$ & Bulk-ESS & Tail-ESS\\
\midrule
\cellcolor{gray!10}{categorical} & \cellcolor{gray!10}{1} & \cellcolor{gray!10}{0.047} & \cellcolor{gray!10}{{}[.040, .056]} & \cellcolor{gray!10}{1.000} & \cellcolor{gray!10}{14618} & \cellcolor{gray!10}{27748}\\
categorical & 3 & 0.052 & {}[.044, .061] & 1.000 & 16317 & 30373\\
\cellcolor{gray!10}{original} & \cellcolor{gray!10}{1} & \cellcolor{gray!10}{0.058} & \cellcolor{gray!10}{{}[.049, .066]} & \cellcolor{gray!10}{1.000} & \cellcolor{gray!10}{17747} & \cellcolor{gray!10}{32562}\\
original & 3 & 0.068 & {}[.060, .077] & 1.000 & 18043 & 31776\\
\cellcolor{gray!10}{recalibrated} & \cellcolor{gray!10}{1} & \cellcolor{gray!10}{0.059} & \cellcolor{gray!10}{{}[.050, .067]} & \cellcolor{gray!10}{1.000} & \cellcolor{gray!10}{17964} & \cellcolor{gray!10}{32200}\\
\addlinespace
recalibrated & 3 & 0.061 & {}[.053, .070] & 1.000 & 18081 & 32889\\
\bottomrule
\end{tabular}
    \caption{Posterior expected value of a participants' \textbf{resolution} (i.e., Brier score decomposition measure of discrimination) for each condition in each phase.}
    \label{tab:resolution_prediction}
\end{table}

\begin{table}[H]
\centering
\begin{tabular}{lrlrrr}
\toprule
Condition & MAP & 95\% HDI & $\hat{R}$ & Bulk-ESS & Tail-ESS\\
\midrule
\cellcolor{gray!10}{categorical} & \cellcolor{gray!10}{0.005} & \cellcolor{gray!10}{{}[-.002, .011]} & \cellcolor{gray!10}{1.000} & \cellcolor{gray!10}{72541} & \cellcolor{gray!10}{53553}\\
original & 0.010 & {}[.004, .017] & 1.000 & 82487 & 55296\\
\cellcolor{gray!10}{recalibrated} & \cellcolor{gray!10}{0.002} & \cellcolor{gray!10}{{}[-.004, .008]} & \cellcolor{gray!10}{1.000} & \cellcolor{gray!10}{84679} & \cellcolor{gray!10}{54921}\\
\bottomrule
\end{tabular}
    \caption{Posterior expected \textbf{change in resolution} for each condition from Phase 1 to 3.}
    \label{tab:resolution_change}
\end{table}

\begin{table}[H]
\centering
\begin{tabular}{lrlrrr}
\toprule
Comparison & MAP & 95\% HDI & $\hat{R}$ & Bulk-ESS & Tail-ESS\\
\midrule
\cellcolor{gray!10}{original - categorical} & \cellcolor{gray!10}{0.006} & \cellcolor{gray!10}{{}[-.003, .014]} & \cellcolor{gray!10}{1.000} & \cellcolor{gray!10}{79050} & \cellcolor{gray!10}{54610}\\
recalibrated - categorical & -0.002 & {}[-.011, .007] & 1.000 & 81954 & 55193\\
\cellcolor{gray!10}{original - recalibrated} & \cellcolor{gray!10}{0.008} & \cellcolor{gray!10}{{}[-.001, .017]} & \cellcolor{gray!10}{1.001} & \cellcolor{gray!10}{113523} & \cellcolor{gray!10}{41582}\\
\bottomrule
\end{tabular}
    \caption{Posterior expected \textbf{difference between conditions in the change in resolution} from Phase 1 to Phase 3.}
    \label{tab:resolution_difinchange}
\end{table}

\begin{table}[H]
\centering
\begin{tabular}{lrlrrr}
\toprule
Parameter & MAP & 95\% HDI & $\hat{R}$ & Bulk-ESS & Tail-ESS\\
\midrule
\addlinespace[0.3em]
\multicolumn{6}{l}{\textbf{Fixed Effects $\beta$s}}\\
\hspace{1em}\cellcolor{gray!10}{Intercept} & \cellcolor{gray!10}{0.047} & \cellcolor{gray!10}{{}[.040, .056]} & \cellcolor{gray!10}{1.000} & \cellcolor{gray!10}{14618} & \cellcolor{gray!10}{27748}\\
\hspace{1em}Cond. orig. (vs cat.) & 0.010 & {}[-.001, .022] & 1.000 & 15649 & 28012\\
\hspace{1em}\cellcolor{gray!10}{Cond. recal. (vs cat.)} & \cellcolor{gray!10}{0.011} & \cellcolor{gray!10}{{}[-.001, .023]} & \cellcolor{gray!10}{1.000} & \cellcolor{gray!10}{15298} & \cellcolor{gray!10}{28817}\\
\hspace{1em}Phase 3 (vs 1) & 0.005 & {}[-.002, .011] & 1.000 & 72541 & 53553\\
\hspace{1em}\cellcolor{gray!10}{Cond. orig. (vs cat.) × Phase 3 (vs 1)} & \cellcolor{gray!10}{0.006} & \cellcolor{gray!10}{{}[-.003, .014]} & \cellcolor{gray!10}{1.000} & \cellcolor{gray!10}{79050} & \cellcolor{gray!10}{54610}\\
\hspace{1em}Cond. recal. (vs cat.) × Phase 3 (vs 1) & -0.002 & {}[-.011, .007] & 1.000 & 81954 & 55193\\
\addlinespace[0.3em]
\multicolumn{6}{l}{\textbf{Residual SD}}\\
\hspace{1em}\cellcolor{gray!10}{Sigma} & \cellcolor{gray!10}{0.023} & \cellcolor{gray!10}{{}[.021, .024]} & \cellcolor{gray!10}{1.000} & \cellcolor{gray!10}{30767} & \cellcolor{gray!10}{43215}\\
\addlinespace[0.3em]
\multicolumn{6}{l}{\textbf{Random Effect SDs}}\\
\addlinespace[0.3em]
\multicolumn{6}{l}{\textbf{\quad Group: participant}}\\
\hspace{1em}\hspace{1em}Intercept & 0.038 & {}[.034, .042] & 1.000 & 17138 & 31836\\
\bottomrule
\end{tabular}
    \caption{Population-level parameters from linear mixed-effects model of participant resolution.}
    \label{tab:resolution_parameters}
\end{table}

\subsection{Binarized accuracy}

\begin{table}[H]
\centering
\begin{tabular}{llrlrrr}
\toprule
Condition & Phase & MAP & 95\% HDI & $\hat{R}$ & Bulk-ESS & Tail-ESS\\
\midrule
\cellcolor{gray!10}{categorical} & \cellcolor{gray!10}{1} & \cellcolor{gray!10}{0.673} & \cellcolor{gray!10}{{}[.648, .700]} & \cellcolor{gray!10}{1.001} & \cellcolor{gray!10}{10797} & \cellcolor{gray!10}{21238}\\
categorical & 3 & 0.699 & {}[.671, .728] & 1.001 & 12897 & 26354\\
\cellcolor{gray!10}{original} & \cellcolor{gray!10}{1} & \cellcolor{gray!10}{0.703} & \cellcolor{gray!10}{{}[.673, .729]} & \cellcolor{gray!10}{1.000} & \cellcolor{gray!10}{12045} & \cellcolor{gray!10}{23355}\\
original & 3 & 0.739 & {}[.712, .765] & 1.000 & 15817 & 29611\\
\cellcolor{gray!10}{recalibrated} & \cellcolor{gray!10}{1} & \cellcolor{gray!10}{0.692} & \cellcolor{gray!10}{{}[.662, .718]} & \cellcolor{gray!10}{1.001} & \cellcolor{gray!10}{10646} & \cellcolor{gray!10}{20899}\\
\addlinespace
recalibrated & 3 & 0.722 & {}[.690, .750] & 1.001 & 12885 & 27559\\
\bottomrule
\end{tabular}
    \caption{Posterior expected value of a participants' \textbf{binarized accuracy} (i.e., trial-level probability of judging a cell to have a $>50\%$ probability of being blast when it is blast or a $<50\%$ probability of being blast when it is non-blast) for each condition in each phase.}
    \label{tab:accuracy_prediction}
\end{table}

\begin{table}[H]
\centering
\begin{tabular}{lrlrrr}
\toprule
Condition & MAP & 95\% HDI & $\hat{R}$ & Bulk-ESS & Tail-ESS\\
\midrule
\cellcolor{gray!10}{categorical} & \cellcolor{gray!10}{0.026} & \cellcolor{gray!10}{{}[.005, .044]} & \cellcolor{gray!10}{1.000} & \cellcolor{gray!10}{27714} & \cellcolor{gray!10}{39887}\\
original & 0.038 & {}[.016, .058] & 1.000 & 19297 & 38747\\
\cellcolor{gray!10}{recalibrated} & \cellcolor{gray!10}{0.030} & \cellcolor{gray!10}{{}[.010, .049]} & \cellcolor{gray!10}{1.000} & \cellcolor{gray!10}{37565} & \cellcolor{gray!10}{47778}\\
\bottomrule
\end{tabular}
    \caption{Posterior expected \textbf{change in binarized accuracy} for each condition from Phase 1 to 3.}
    \label{tab:accuracy_change}
\end{table}

\begin{table}[H]
\centering
\begin{tabular}{lrlrrr}
\toprule
Comparison & MAP & 95\% HDI & $\hat{R}$ & Bulk-ESS & Tail-ESS\\
\midrule
\cellcolor{gray!10}{original - categorical} & \cellcolor{gray!10}{0.011} & \cellcolor{gray!10}{{}[-.015, .040]} & \cellcolor{gray!10}{1.000} & \cellcolor{gray!10}{21378} & \cellcolor{gray!10}{38083}\\
recalibrated - categorical & 0.005 & {}[-.021, .030] & 1.000 & 34499 & 45348\\
\cellcolor{gray!10}{original - recalibrated} & \cellcolor{gray!10}{0.007} & \cellcolor{gray!10}{{}[-.020, .034]} & \cellcolor{gray!10}{1.000} & \cellcolor{gray!10}{34359} & \cellcolor{gray!10}{48932}\\
\bottomrule
\end{tabular}
    \caption{Posterior expected \textbf{difference between conditions in the change in accuracy} from Phase 1 to Phase 3.}
    \label{tab:accuracy_difinchange}
\end{table}

\begin{longtable}[H]{lrlrrr}
\toprule
Parameter & MAP & 95\% HDI & $\hat{R}$ & Bulk-ESS & Tail-ESS\\
\midrule
\addlinespace[0.3em]
\multicolumn{6}{l}{\textbf{Fixed Effects $\beta$s}}\\
\hspace{1em}\cellcolor{gray!10}{Intercept} & \cellcolor{gray!10}{0.722} & \cellcolor{gray!10}{{}[.609, .845]} & \cellcolor{gray!10}{1.001} & \cellcolor{gray!10}{10797} & \cellcolor{gray!10}{21238}\\
\hspace{1em}Cond. orig. (vs cat.) & 0.127 & {}[-.031, .282] & 1.001 & 11563 & 22656\\
\hspace{1em}\cellcolor{gray!10}{Cond. recal. (vs cat.)} & \cellcolor{gray!10}{0.078} & \cellcolor{gray!10}{{}[-.081, .230]} & \cellcolor{gray!10}{1.000} & \cellcolor{gray!10}{11688} & \cellcolor{gray!10}{23313}\\
\hspace{1em}Phase 3 (vs 1) & 0.110 & {}[.024, .208] & 1.000 & 28071 & 40544\\
\hspace{1em}\cellcolor{gray!10}{Cond. orig. (vs cat.) × Phase 3 (vs 1)} & \cellcolor{gray!10}{0.063} & \cellcolor{gray!10}{{}[-.067, .202]} & \cellcolor{gray!10}{1.000} & \cellcolor{gray!10}{21763} & \cellcolor{gray!10}{40318}\\
\hspace{1em}Cond. recal. (vs cat.) × Phase 3 (vs 1) & 0.028 & {}[-.097, .149] & 1.000 & 34997 & 45671\\
\addlinespace[0.3em]
\multicolumn{6}{l}{\textbf{Random Effect SDs}}\\
\addlinespace[0.3em]
\multicolumn{6}{l}{\textbf{\quad Group: stimulus (blast)}}\\
\hspace{1em}\hspace{1em}\cellcolor{gray!10}{Intercept} & \cellcolor{gray!10}{0.366} & \cellcolor{gray!10}{{}[.311, .433]} & \cellcolor{gray!10}{1.000} & \cellcolor{gray!10}{21790} & \cellcolor{gray!10}{38839}\\
\hspace{1em}\hspace{1em}Cond. orig. (vs cat.) & 0.140 & {}[.001, .230] & 1.001 & 9685 & 15725\\
\hspace{1em}\hspace{1em}\cellcolor{gray!10}{Cond. recal. (vs cat.)} & \cellcolor{gray!10}{0.002} & \cellcolor{gray!10}{{}[.000, .115]} & \cellcolor{gray!10}{1.000} & \cellcolor{gray!10}{19070} & \cellcolor{gray!10}{29723}\\
\hspace{1em}\hspace{1em}Phase 3 (vs 1) & 0.007 & {}[.000, .155] & 1.001 & 8381 & 18297\\
\hspace{1em}\hspace{1em}\cellcolor{gray!10}{Cond. orig. (vs cat.) × Phase 3 (vs 1)} & \cellcolor{gray!10}{0.181} & \cellcolor{gray!10}{{}[.000, .305]} & \cellcolor{gray!10}{1.001} & \cellcolor{gray!10}{7134} & \cellcolor{gray!10}{12317}\\
\hspace{1em}\hspace{1em}Cond. recal. (vs cat.) × Phase 3 (vs 1) & 0.121 & {}[.000, .251] & 1.001 & 9468 & 20040\\
\addlinespace[0.3em]
\multicolumn{6}{l}{\textbf{\quad Group: stimulus (non-blast)}}\\
\hspace{1em}\hspace{1em}\cellcolor{gray!10}{Intercept} & \cellcolor{gray!10}{0.702} & \cellcolor{gray!10}{{}[.612, .806]} & \cellcolor{gray!10}{1.000} & \cellcolor{gray!10}{14577} & \cellcolor{gray!10}{28710}\\
\hspace{1em}\hspace{1em}Cond. orig. (vs cat.) & 0.158 & {}[.000, .279] & 1.002 & 5695 & 11412\\
\hspace{1em}\hspace{1em}\cellcolor{gray!10}{Cond. recal. (vs cat.)} & \cellcolor{gray!10}{0.016} & \cellcolor{gray!10}{{}[.000, .172]} & \cellcolor{gray!10}{1.001} & \cellcolor{gray!10}{5864} & \cellcolor{gray!10}{17647}\\
\hspace{1em}\hspace{1em}Phase 3 (vs 1) & 0.281 & {}[.194, .374] & 1.000 & 14145 & 31243\\
\hspace{1em}\hspace{1em}\cellcolor{gray!10}{Cond. orig. (vs cat.) × Phase 3 (vs 1)} & \cellcolor{gray!10}{0.197} & \cellcolor{gray!10}{{}[.000, .383]} & \cellcolor{gray!10}{1.001} & \cellcolor{gray!10}{5484} & \cellcolor{gray!10}{14688}\\
\hspace{1em}\hspace{1em}Cond. recal. (vs cat.) × Phase 3 (vs 1) & 0.000 & {}[.000, .203] & 1.001 & 16515 & 29000\\
\addlinespace[0.3em]
\multicolumn{6}{l}{\textbf{\quad Group: participant (in categorical AI condition)}}\\
\hspace{1em}\hspace{1em}\cellcolor{gray!10}{Intercept} & \cellcolor{gray!10}{0.485} & \cellcolor{gray!10}{{}[.408, .578]} & \cellcolor{gray!10}{1.000} & \cellcolor{gray!10}{22248} & \cellcolor{gray!10}{37146}\\
\hspace{1em}\hspace{1em}Phase 3 (vs 1) & 0.311 & {}[.224, .400] & 1.000 & 20781 & 32329\\
\addlinespace[0.3em]
\multicolumn{6}{l}{\textbf{\quad Group: participant (in original AI condition)}}\\
\hspace{1em}\hspace{1em}\cellcolor{gray!10}{Intercept} & \cellcolor{gray!10}{0.562} & \cellcolor{gray!10}{{}[.479, .666]} & \cellcolor{gray!10}{1.000} & \cellcolor{gray!10}{18481} & \cellcolor{gray!10}{34549}\\
\hspace{1em}\hspace{1em}Phase 3 (vs 1) & 0.307 & {}[.218, .401] & 1.000 & 20897 & 31982\\
\addlinespace[0.3em]
\multicolumn{6}{l}{\textbf{\quad Group: participant (in recalibrated AI condition)}}\\
\hspace{1em}\hspace{1em}\cellcolor{gray!10}{Intercept} & \cellcolor{gray!10}{0.572} & \cellcolor{gray!10}{{}[.487, .677]} & \cellcolor{gray!10}{1.000} & \cellcolor{gray!10}{17632} & \cellcolor{gray!10}{31744}\\
\hspace{1em}\hspace{1em}Phase 3 (vs 1) & 0.289 & {}[.198, .383] & 1.000 & 18399 & 28135\\
\bottomrule
    \caption{Population-level parameters from logistic mixed-effects model of participant probability of correct response, binarized about 50\%.}
    \label{tab:accuracy_parameters}
\end{longtable}

Note: of the correlations between random effects, the following were credibly nonzero, or close to such:

\begin{itemize}
    \item The random intercept of non-blast stimuli and the random effect of phase being Phase 3 had a positive correlation (MAP = 0.626 (95\% HDI = [0.312, 0.856])). This means that non-blast stimuli which were more likely to be correctly classified in Phase 1 (the reference phase) also saw a greater improvement in likelihood of correct classification by Phase 3
    \item The random intercept of a subject in the original AI condition and the random effect of phase being Phase 3 had some evidence, though inconclusive, of a negative correlation (MAP = -.246 (95\% HDI = [-.505, .061])). This means that there is some evidence that original condition participants who were better in Phase 1 saw less improvement (or more deterioration) by Phase 3. Put another way, this would mean that greater improvement tended to came from those with lesser initial performance.
\end{itemize}

\subsection{Expected Calibration Error (ECE)}

\begin{table}[H]
\centering
\begin{tabular}{llrlrrr}
\toprule
Condition & Phase & MAP & 95\% HDI & $\hat{R}$ & Bulk-ESS & Tail-ESS\\
\midrule
\cellcolor{gray!10}{categorical} & \cellcolor{gray!10}{1} & \cellcolor{gray!10}{0.214} & \cellcolor{gray!10}{{}[.195, .233]} & \cellcolor{gray!10}{1.000} & \cellcolor{gray!10}{30675} & \cellcolor{gray!10}{40121}\\
categorical & 3 & 0.215 & {}[.195, .236] & 1.000 & 34907 & 42787\\
\cellcolor{gray!10}{original} & \cellcolor{gray!10}{1} & \cellcolor{gray!10}{0.206} & \cellcolor{gray!10}{{}[.186, .226]} & \cellcolor{gray!10}{1.000} & \cellcolor{gray!10}{36432} & \cellcolor{gray!10}{44836}\\
original & 3 & 0.159 & {}[.139, .180] & 1.000 & 36258 & 46536\\
\cellcolor{gray!10}{recalibrated} & \cellcolor{gray!10}{1} & \cellcolor{gray!10}{0.208} & \cellcolor{gray!10}{{}[.189, .228]} & \cellcolor{gray!10}{1.000} & \cellcolor{gray!10}{37470} & \cellcolor{gray!10}{48716}\\
\addlinespace
recalibrated & 3 & 0.216 & {}[.195, .236] & 1.000 & 38595 & 49492\\
\bottomrule
\end{tabular}
    \caption{Posterior expected value of a participants' \textbf{expected calibration error (ECE)} (i.e., a measure of the average absolute calibration error) for each condition in each phase.}
    \label{tab:ece_prediction}
\end{table}

\begin{table}[H]
\centering
\begin{tabular}{lrlrrr}
\toprule
Condition & MAP & 95\% HDI & $\hat{R}$ & Bulk-ESS & Tail-ESS\\
\midrule
\cellcolor{gray!10}{categorical} & \cellcolor{gray!10}{0.001} & \cellcolor{gray!10}{{}[-.015, .018]} & \cellcolor{gray!10}{1.000} & \cellcolor{gray!10}{71913} & \cellcolor{gray!10}{50778}\\
original & -0.046 & {}[-.063, -.030] & 1.000 & 79664 & 53360\\
\cellcolor{gray!10}{recalibrated} & \cellcolor{gray!10}{0.007} & \cellcolor{gray!10}{{}[-.010, .024]} & \cellcolor{gray!10}{1.000} & \cellcolor{gray!10}{82407} & \cellcolor{gray!10}{53217}\\
\bottomrule
\end{tabular}
    \caption{Posterior expected \textbf{change in ECE} for each condition from Phase 1 to 3.}
    \label{tab:ece_change}
\end{table}

\begin{table}[H]
\centering
\begin{tabular}{lrlrrr}
\toprule
Comparison & MAP & 95\% HDI & $\hat{R}$ & Bulk-ESS & Tail-ESS\\
\midrule
\cellcolor{gray!10}{original - categorical} & \cellcolor{gray!10}{-0.048} & \cellcolor{gray!10}{{}[-.072, -.025]} & \cellcolor{gray!10}{1.000} & \cellcolor{gray!10}{75313} & \cellcolor{gray!10}{54364}\\
recalibrated - categorical & 0.006 & {}[-.017, .029] & 1.000 & 77471 & 55129\\
\cellcolor{gray!10}{original - recalibrated} & \cellcolor{gray!10}{-0.055} & \cellcolor{gray!10}{{}[-.077, -.030]} & \cellcolor{gray!10}{1.000} & \cellcolor{gray!10}{95632} & \cellcolor{gray!10}{43835}\\
\bottomrule
\end{tabular}
    \caption{Posterior expected \textbf{difference between conditions in the change in ECE} from Phase 1 to Phase 3.}
    \label{tab:ece_difinchange}
\end{table}

\begin{table}[H]
\centering
\begin{tabular}{lrlrrr}
\toprule
Parameter & MAP & 95\% HDI & $\hat{R}$ & Bulk-ESS & Tail-ESS\\
\midrule
\addlinespace[0.3em]
\multicolumn{6}{l}{\textbf{Fixed Effects $\beta$s}}\\
\hspace{1em}\cellcolor{gray!10}{Intercept} & \cellcolor{gray!10}{0.214} & \cellcolor{gray!10}{{}[.195, .233]} & \cellcolor{gray!10}{1.000} & \cellcolor{gray!10}{30675} & \cellcolor{gray!10}{40121}\\
\hspace{1em}Cond. orig. (vs cat.) & -0.007 & {}[-.033, .017] & 1.000 & 33855 & 42155\\
\hspace{1em}\cellcolor{gray!10}{Cond. recal. (vs cat.)} & \cellcolor{gray!10}{-0.006} & \cellcolor{gray!10}{{}[-.031, .019]} & \cellcolor{gray!10}{1.000} & \cellcolor{gray!10}{35015} & \cellcolor{gray!10}{44746}\\
\hspace{1em}Phase 3 (vs 1) & 0.001 & {}[-.015, .018] & 1.000 & 71913 & 50778\\
\hspace{1em}\cellcolor{gray!10}{Cond. orig. (vs cat.) × Phase 3 (vs 1)} & \cellcolor{gray!10}{-0.048} & \cellcolor{gray!10}{{}[-.072, -.025]} & \cellcolor{gray!10}{1.000} & \cellcolor{gray!10}{75313} & \cellcolor{gray!10}{54364}\\
\hspace{1em}Cond. recal. (vs cat.) × Phase 3 (vs 1) & 0.006 & {}[-.017, .029] & 1.000 & 77471 & 55129\\
\addlinespace[0.3em]
\multicolumn{6}{l}{\textbf{Residual SD}}\\
\hspace{1em}\cellcolor{gray!10}{Sigma} & \cellcolor{gray!10}{0.061} & \cellcolor{gray!10}{{}[.057, .067]} & \cellcolor{gray!10}{1.000} & \cellcolor{gray!10}{29442} & \cellcolor{gray!10}{43152}\\
\addlinespace[0.3em]
\multicolumn{6}{l}{\textbf{Random Effect SDs}}\\
\addlinespace[0.3em]
\multicolumn{6}{l}{\textbf{\quad Group: participant}}\\
\hspace{1em}\hspace{1em}Intercept & 0.090 & {}[.081, .099] & 1.000 & 20880 & 33985\\
\bottomrule
\end{tabular}
    \caption{Population-level parameters from linear mixed-effects model of participant expected calibration error (ECE).}
    \label{tab:ece_parameters}
\end{table}

\section{Linear in Log Odds (LLO) model}

\subsection{Over-/under-extremity ($\alpha$)}

\begin{table}[H]
\centering
\begin{tabular}{llrlrrr}
\toprule
Condition & Phase & MAP & 95\% HDI & $\hat{R}$ & Bulk-ESS & Tail-ESS\\
\midrule
\cellcolor{gray!10}{categorical} & \cellcolor{gray!10}{1} & \cellcolor{gray!10}{0.285} & \cellcolor{gray!10}{{}[.224, .355]} & \cellcolor{gray!10}{1.000} & \cellcolor{gray!10}{19583} & \cellcolor{gray!10}{31561}\\
categorical & 3 & 0.333 & {}[.257, .419] & 1.000 & 19062 & 30615\\
\cellcolor{gray!10}{original} & \cellcolor{gray!10}{1} & \cellcolor{gray!10}{0.291} & \cellcolor{gray!10}{{}[.235, .344]} & \cellcolor{gray!10}{1.000} & \cellcolor{gray!10}{40871} & \cellcolor{gray!10}{43783}\\
original & 3 & 0.547 & {}[.457, .644] & 1.000 & 37083 & 42475\\
\cellcolor{gray!10}{recalibrated} & \cellcolor{gray!10}{1} & \cellcolor{gray!10}{0.275} & \cellcolor{gray!10}{{}[.224, .326]} & \cellcolor{gray!10}{1.000} & \cellcolor{gray!10}{46084} & \cellcolor{gray!10}{45363}\\
\addlinespace
recalibrated & 3 & 0.276 & {}[.227, .325] & 1.000 & 50027 & 48149\\
\bottomrule
\end{tabular}
    \caption{Posterior expected value of a participants' \textbf{over-/under-extremity ($\alpha$)} for each condition in each phase. Values of $\alpha < 1$ indicate over-extremity, with $\alpha=0$ indicating that judgments have no diagnostic value. Values of $\alpha>1$ indicate under-extremity of judgments.}
    \label{tab:llo.alpha_prediction}
\end{table}

\begin{table}[H]
\centering
\begin{tabular}{lrlrrr}
\toprule
Condition & MAP & 95\% HDI & $\hat{R}$ & Bulk-ESS & Tail-ESS\\
\midrule
\cellcolor{gray!10}{categorical} & \cellcolor{gray!10}{0.045} & \cellcolor{gray!10}{{}[.011, .085]} & \cellcolor{gray!10}{1.000} & \cellcolor{gray!10}{30856} & \cellcolor{gray!10}{37119}\\
original & 0.260 & {}[.195, .327] & 1.000 & 42106 & 44870\\
\cellcolor{gray!10}{recalibrated} & \cellcolor{gray!10}{0.002} & \cellcolor{gray!10}{{}[-.030, .034]} & \cellcolor{gray!10}{1.000} & \cellcolor{gray!10}{54429} & \cellcolor{gray!10}{53145}\\
\bottomrule
\end{tabular}
    \caption{Posterior expected \textbf{change in $\alpha$} for each condition from Phase 1 to 3.}
    \label{tab:llo.alpha_change}
\end{table}

\begin{table}[H]
\centering
\begin{tabular}{lrlrrr}
\toprule
Comparison & MAP & 95\% HDI & $\hat{R}$ & Bulk-ESS & Tail-ESS\\
\midrule
\cellcolor{gray!10}{original - categorical} & \cellcolor{gray!10}{0.214} & \cellcolor{gray!10}{{}[.137, .289]} & \cellcolor{gray!10}{1.000} & \cellcolor{gray!10}{34117} & \cellcolor{gray!10}{41982}\\
recalibrated - categorical & -0.044 & {}[-.096, .002] & 1.000 & 33719 & 41698\\
\cellcolor{gray!10}{original - recalibrated} & \cellcolor{gray!10}{0.256} & \cellcolor{gray!10}{{}[.186, .333]} & \cellcolor{gray!10}{1.000} & \cellcolor{gray!10}{44296} & \cellcolor{gray!10}{47017}\\
\bottomrule
\end{tabular}
    \caption{Posterior expected \textbf{difference between conditions in the change in $\alpha$} from Phase 1 to Phase 3.}
    \label{tab:llo.alpha_difinchange}
\end{table}

\subsection{Over-/under-prediction ($\beta$)}

\begin{table}[H]
\centering
\begin{tabular}{llrlrrr}
\toprule
Condition & Phase & MAP & 95\% HDI & $\hat{R}$ & Bulk-ESS & Tail-ESS\\
\midrule
\cellcolor{gray!10}{categorical} & \cellcolor{gray!10}{1} & \cellcolor{gray!10}{0.069} & \cellcolor{gray!10}{{}[.014, .117]} & \cellcolor{gray!10}{1.000} & \cellcolor{gray!10}{32350} & \cellcolor{gray!10}{43608}\\
categorical & 3 & 0.023 & {}[-.029, .076] & 1.000 & 55751 & 52433\\
\cellcolor{gray!10}{original} & \cellcolor{gray!10}{1} & \cellcolor{gray!10}{0.126} & \cellcolor{gray!10}{{}[.075, .181]} & \cellcolor{gray!10}{1.000} & \cellcolor{gray!10}{58707} & \cellcolor{gray!10}{53305}\\
original & 3 & -0.034 & {}[-.102, .032] & 1.000 & 46595 & 51762\\
\cellcolor{gray!10}{recalibrated} & \cellcolor{gray!10}{1} & \cellcolor{gray!10}{0.076} & \cellcolor{gray!10}{{}[.020, .134]} & \cellcolor{gray!10}{1.000} & \cellcolor{gray!10}{50872} & \cellcolor{gray!10}{54046}\\
\addlinespace
recalibrated & 3 & -0.007 & {}[-.068, .054] & 1.000 & 53618 & 56207\\
\bottomrule
\end{tabular}
    \caption{Posterior expected value of a participants' \textbf{over-/under-prediction ($\beta$)} for each condition in each phase. Values of $\beta < 0$ indicate over-prediction, where subjective judgments of the probability of cells being blast are too high with respect to corresponding relative frequencies. Values of $\beta > 0$ indicate under-prediction, where subjective probabilities are too low.}
    \label{tab:llo.beta_prediction}
\end{table}

\begin{table}[H]
\centering
\begin{tabular}{lrlrrr}
\toprule
Condition & MAP & 95\% HDI & $\hat{R}$ & Bulk-ESS & Tail-ESS\\
\midrule
\cellcolor{gray!10}{categorical} & \cellcolor{gray!10}{-0.040} & \cellcolor{gray!10}{{}[-.103, .018]} & \cellcolor{gray!10}{1.000} & \cellcolor{gray!10}{40633} & \cellcolor{gray!10}{48023}\\
original & -0.161 & {}[-.238, -.086] & 1.000 & 52322 & 54858\\
\cellcolor{gray!10}{recalibrated} & \cellcolor{gray!10}{-0.084} & \cellcolor{gray!10}{{}[-.150, -.017]} & \cellcolor{gray!10}{1.000} & \cellcolor{gray!10}{64345} & \cellcolor{gray!10}{55299}\\
\bottomrule
\end{tabular}
    \caption{Posterior expected \textbf{change in $\beta$} for each condition from Phase 1 to 3.}
    \label{tab:llo.beta_change}
\end{table}

\begin{table}[H]
\centering
\begin{tabular}{lrlrrr}
\toprule
Comparison & MAP & 95\% HDI & $\hat{R}$ & Bulk-ESS & Tail-ESS\\
\midrule
\cellcolor{gray!10}{original - categorical} & \cellcolor{gray!10}{-0.125} & \cellcolor{gray!10}{{}[-.217, -.022]} & \cellcolor{gray!10}{1.000} & \cellcolor{gray!10}{40466} & \cellcolor{gray!10}{48028}\\
recalibrated - categorical & -0.039 & {}[-.130, .049] & 1.000 & 41768 & 50125\\
\cellcolor{gray!10}{original - recalibrated} & \cellcolor{gray!10}{-0.076} & \cellcolor{gray!10}{{}[-.182, .021]} & \cellcolor{gray!10}{1.000} & \cellcolor{gray!10}{62861} & \cellcolor{gray!10}{52167}\\
\bottomrule
\end{tabular}
    \caption{Posterior expected \textbf{difference between conditions in the change in $\beta$} from Phase 1 to Phase 3.}
    \label{tab:llo.beta_difinchange}
\end{table}

\subsection{Model in GLM parameterization}

\begin{longtable}[H]{lrlrrr}
\toprule
Parameter & MAP & 95\% HDI & $\hat{R}$ & Bulk-ESS & Tail-ESS\\
\midrule
\addlinespace[0.3em]
\multicolumn{6}{l}{\textbf{Fixed Effect $\beta$s}}\\
\hspace{1em}\cellcolor{gray!10}{Intercept} & \cellcolor{gray!10}{0.069} & \cellcolor{gray!10}{{}[.014, .117]} & \cellcolor{gray!10}{1.000} & \cellcolor{gray!10}{32350} & \cellcolor{gray!10}{43608}\\
\hspace{1em}Cond. orig. (vs cat.) & 0.057 & {}[-.013, .136] & 1.000 & 34491 & 44656\\
\hspace{1em}\cellcolor{gray!10}{Cond. recal. (vs cat.)} & \cellcolor{gray!10}{0.010} & \cellcolor{gray!10}{{}[-.067, .087]} & \cellcolor{gray!10}{1.000} & \cellcolor{gray!10}{33444} & \cellcolor{gray!10}{45235}\\
\hspace{1em}Logit judgment & 0.285 & {}[.224, .355] & 1.000 & 19583 & 31561\\
\hspace{1em}\cellcolor{gray!10}{Phase 3 (vs 1)} & \cellcolor{gray!10}{-0.040} & \cellcolor{gray!10}{{}[-.103, .018]} & \cellcolor{gray!10}{1.000} & \cellcolor{gray!10}{40634} & \cellcolor{gray!10}{48023}\\
\hspace{1em}Cond. orig. (vs cat.) × Logit judgment & -0.003 & {}[-.084, .086] & 1.000 & 21811 & 33980\\
\hspace{1em}\cellcolor{gray!10}{Cond. orig. (vs cat.) × Phase 3 (vs 1)} & \cellcolor{gray!10}{-0.125} & \cellcolor{gray!10}{{}[-.217, -.022]} & \cellcolor{gray!10}{1.000} & \cellcolor{gray!10}{40467} & \cellcolor{gray!10}{48028}\\
\hspace{1em}Cond. recal. (vs cat.) × Logit judgment & -0.018 & {}[-.097, .070] & 1.000 & 22801 & 35082\\
\hspace{1em}\cellcolor{gray!10}{Cond. recal. (vs cat.) × Phase 3 (vs 1)} & \cellcolor{gray!10}{-0.039} & \cellcolor{gray!10}{{}[-.130, .049]} & \cellcolor{gray!10}{1.000} & \cellcolor{gray!10}{41768} & \cellcolor{gray!10}{50125}\\
\hspace{1em}Phase 3 (vs 1) × Logit judgment & 0.045 & {}[.011, .085] & 1.000 & 30856 & 37119\\
\hspace{1em}\cellcolor{gray!10}{\makecell[l]{Cond. orig. (vs cat.) × Phase 3 (vs 1)\\\quad× Logit judgment}} & \cellcolor{gray!10}{0.214} & \cellcolor{gray!10}{{}[.137, .289]} & \cellcolor{gray!10}{1.000} & \cellcolor{gray!10}{34117} & \cellcolor{gray!10}{41982}\\
\hspace{1em}\makecell[l]{Cond. recal. (vs cat.) × Phase 3 (vs 1)\\\quad× Logit judgment} & -0.044 & {}[-.096, .002] & 1.000 & 33719 & 41698\\
\addlinespace[0.3em]
\multicolumn{6}{l}{\textbf{Random Effect SDs}}\\
\addlinespace[0.3em]
\multicolumn{6}{l}{\textbf{\quad Group: participant (in categorical AI condition)}}\\
\hspace{1em}\hspace{1em}\cellcolor{gray!10}{Intercept} & \cellcolor{gray!10}{0.142} & \cellcolor{gray!10}{{}[.081, .197]} & \cellcolor{gray!10}{1.001} & \cellcolor{gray!10}{19213} & \cellcolor{gray!10}{24832}\\
\hspace{1em}\hspace{1em}Logit judgment & 0.314 & {}[.258, .379] & 1.000 & 25663 & 42156\\
\hspace{1em}\hspace{1em}\cellcolor{gray!10}{Phase 3 (vs 1)} & \cellcolor{gray!10}{0.002} & \cellcolor{gray!10}{{}[.000, .111]} & \cellcolor{gray!10}{1.001} & \cellcolor{gray!10}{11709} & \cellcolor{gray!10}{30004}\\
\hspace{1em}\hspace{1em}Phase 3 (vs 1) × Logit judgment & 0.110 & {}[.068, .162] & 1.000 & 27246 & 37479\\
\addlinespace[0.3em]
\multicolumn{6}{l}{\textbf{\quad Group: participant (in original AI condition)}}\\
\hspace{1em}\hspace{1em}\cellcolor{gray!10}{Intercept} & \cellcolor{gray!10}{0.137} & \cellcolor{gray!10}{{}[.076, .198]} & \cellcolor{gray!10}{1.000} & \cellcolor{gray!10}{27706} & \cellcolor{gray!10}{33786}\\
\hspace{1em}\hspace{1em}Logit judgment & 0.253 & {}[.206, .311] & 1.000 & 25807 & 41085\\
\hspace{1em}\hspace{1em}\cellcolor{gray!10}{Phase 3 (vs 1)} & \cellcolor{gray!10}{0.204} & \cellcolor{gray!10}{{}[.119, .295]} & \cellcolor{gray!10}{1.001} & \cellcolor{gray!10}{11640} & \cellcolor{gray!10}{25464}\\
\hspace{1em}\hspace{1em}Phase 3 (vs 1) × Logit judgment & 0.263 & {}[.214, .331] & 1.000 & 35441 & 45976\\
\addlinespace[0.3em]
\multicolumn{6}{l}{\textbf{\quad Group: participant (in recalibrated AI condition)}}\\
\hspace{1em}\hspace{1em}\cellcolor{gray!10}{Intercept} & \cellcolor{gray!10}{0.172} & \cellcolor{gray!10}{{}[.109, .237]} & \cellcolor{gray!10}{1.000} & \cellcolor{gray!10}{26271} & \cellcolor{gray!10}{34149}\\
\hspace{1em}\hspace{1em}Logit judgment & 0.234 & {}[.192, .280] & 1.000 & 29230 & 44042\\
\hspace{1em}\hspace{1em}\cellcolor{gray!10}{Phase 3 (vs 1)} & \cellcolor{gray!10}{0.109} & \cellcolor{gray!10}{{}[.008, .195]} & \cellcolor{gray!10}{1.002} & \cellcolor{gray!10}{4978} & \cellcolor{gray!10}{3554}\\
\hspace{1em}\hspace{1em}Phase 3 (vs 1) × Logit judgment & 0.096 & {}[.066, .137] & 1.000 & 18958 & 32672\\
\bottomrule
    \caption{Population-level parameters from logistic mixed-effects LLO model.}
    \label{tab:llo_parameters}
\end{longtable}

\section{Extended Methods}
\label{sec:extended_methods}

\subsection{Experimental Methods}

\subsubsection{Stimulus classification and typicality ratings}

For the full 549-image set, three hematopathology faculty from the Department of Pathology, Microbiology and Immunology at Vanderbilt University Medical Center independently classified each image as ``blast'' or ``non-blast'' and rated how typical each image was of its type on a scale from 1 to 5. The 549-image set includes only images for which all three experts agreed on the blast/non-blast classification. This consensus classification was used as the ground truth cell type. The average of the expert typicality ratings was used to assign images to five difficulty categories: very easy, easy, medium, hard, and very hard \citep{Epping.et.al.2026_ImprovingCrowdsourcingAI}.

\subsubsection{Familiarization and task introduction}

Because participants were generally expected to be novices without previous experience with white blood cell morphology, they began by observing 40 examples of white blood cell images (20 blast, 20 non-blast) after an overview of the study. For each participant, examples were pseudo-randomly sampled without replacement in proportion to the overall distribution of difficulty, and were not seen again for the duration of the experiment. Participants saw 4 examples at a time, 2 non-blast on the left and 2 blast on the right, and could observe each set for as long as desired. Next, participants were introduced to the task of assigning subjective probability judgments to each cell, corresponding to their degree of certainty in the cell being blast or non-blast. At this time, we also qualitatively explained how their chance of winning the \$4.00 bonus would be determined by performance in Phase 3, such that judgments with greater certainty would increase chances if they were correct, but would dramatically worsen chances if they were wrong.

\subsubsection{Trial structure and instructions}

In each of the 100 Phase 1 trials, participants were shown the image of the cell in the center of the screen, below which they provided a number between 0\% and 100\% in a text box, answering the question ``What is the probability the cell is a Blast?''. After responding, participants were shown the true cell type (e.g., ``True cell type: \textbf{Non-Blast}''), while simultaneously being shown their response and the cell image. There was no constraint on the time an individual could take to submit their response, or to observe the feedback screen. After clicking ``Next'', the next trial began. After 50 trials, participants saw a break screen where they were informed of the probability they \textit{would have} had to win the \$4.00 if this had been Phase 3 (as determined by their average Brier score). This chance was recalculated for the second set of 50 trials and shown again at the end of Phase 1, providing a kind of aggregate performance feedback in addition to the trial-by-trial feedback.

Before Phase 2, participants were introduced to the AI model, and told that the AI had been ``trained to classify cells as blast or non-blast'' and ``has an accuracy of about 94\% on images of cells similar to the ones you will see''. We told the participants that after each judgment in Phase 2, they would see the AI's predictions instead of the true cell type, suggesting (but not explicitly directing) ``You may use this information to help as you learn to make better probability judgments of your own''. After making each judgment, participants were shown the image, their own judgment, and the AI prediction as either a probability rounded to the nearest percent (e.g., ``AI prediction: 86\% probability of being a Blast'') or a classification (e.g., ``AI prediction: Blast''), depending on the condition.

In Phase 3, participants were given a chance to take a break after every 45 trials, without receiving feedback on their chance to win the bonus.

\subsubsection{Post-experiment survey}

After completing Phase 3, participants answered a survey about the perceptual attributes they used to make their judgments, the weight of these attributes in the judgment process, and whether the AI exposure helped them learn how to distinguish blast from non-blast. Some of these questions were open-ended, while others systematically investigated the use of specific morphological attributes defined in previous literature \citep{Tsutsui.et.al.2023_WBCAttWhiteBlood}. Participants also answered questions about their everyday use of AI, and demographics. These questions were included for exploratory purposes and are not analyzed in this paper. Finally, participants were told whether they won the \$4.00 bonus and given the opportunity to view the specifics of the bonus probability calculation via Brier score.

\subsection{Analytical Methods}

\subsubsection{MCMC convergence diagnostics}

Across all models, all parameter and estimand $\hat{R} < 1.002$. For parameters, the minimum bulk-ESS was $5{,}483$ and the minimum tail-ESS was $9{,}146$. For fixed effect estimands, the minimum bulk-ESS was $10{,}646$ and the minimum tail-ESS was $20{,}899$.

\subsubsection{Interpretation of reliability and resolution}

Perfect calibration would result in reliability equal to 0, achieved by an individual providing subjective probabilities which perfectly match the corresponding relative frequencies of blast cells among those judgments. Perfect reliability can arise in multiple ways, including always giving correct 0\% or 100\% judgments, always responding 50\% when blasts occur on 50\% of trials, or providing well-calibrated probabilities between these extremes. A score of 1 is the worst possible value and occurs when judgments are completely certain but always wrong, such as assigning every blast cell a 0\% probability of being a blast.

For resolution, suppose that $\bar{y} = .5$ as it is in our experiment. Despite perfect calibration, an individual responding 50\% for every cell earns a resolution of 0, the minimum possible value, reflecting that their judgments provide no useful information beyond the overall frequency. At best, resolution is equal to the \textit{uncertainty}, or variance of the true cell type, which is $.5(1-.5)=.25$ for our experiment. This is earned when blast and non-blast cells are judged with mutually exclusive probabilities. One could achieve this in conjunction with perfect reliability by assigning all blast cells 100\% and all non-blast cells 0\%. However, they could also achieve this by assigning every blast cell a probability $>50\%$ and every non-blast cell a probability $<50\%$, or by answering every cell incorrectly but with complete confidence---though this comes at a cost to reliability.

\subsubsection{Binarized accuracy: random-effects structure}

For the binarized accuracy model, we allowed intercepts and effects of phase, cell type, and their interaction to vary by participant, and for these effects to be correlated among participants. We also allowed intercepts and effects of condition, phase, and their interaction to vary by stimulus, and to be correlated among stimuli, capturing systematic differences in stimulus difficulty and learnability within different conditions.

\subsubsection{LLO model: limiting values, calibration curve, and random effects}

As $\alpha$ increases above 1, probability judgments become increasingly under-extreme. As $\alpha \rightarrow 0$, judgments become increasingly over-extreme, until $\alpha = 0$ where they become entirely undiagnostic of the true category of a cell, corresponding to resolution of 0. Below 0, $\alpha$ indicates that judgments are diagnostically inverted, with non-blast cells getting higher subjective probabilities of being blast, and vice versa.

One way to think of the LLO is as a model-based calibration curve, a commonly used visualization of probability calibration errors. Solving the linear form for $P_\text{LLO}(y_{j}=1 \mid x_j)$ gives the probability that a cell is a blast as a function of subjective probability judgments:
\begin{equation}\label{LLO_pblasteq_supp}
P_\text{LLO}\left(y_j=1 \mid x_j\right) = \frac{1}{1 + e^{-\left(\alpha \,\text{logit}\left(x_{j}\right) + \beta\right)}}
\end{equation}
Plotting this function alongside actual probability judgments shows how the LLO model is capable of capturing systematic errors.

For the LLO model, the value of $\beta$ for each condition and phase is obtained from the overall intercept and the main effects of condition and phase along with their interaction, while the value of $\alpha$ is obtained from all terms interacting with the logit judgment predictor. We allowed baseline $\alpha$ and $\beta$ and their change from Phase 1 to Phase 3 to vary among participants via by-participant varying intercepts, effects of phase and logit judgment, their interaction, and correlations among these effects.

\section{Recalibrated Elicited Beliefs and the Blackwell Ordering}
\label{sec:blackwell_supp}

This section provides the conceptual background for the Blackwell informativeness analysis reported in the main text. We first motivate the distribution of recalibrated elicited beliefs as the object of interest, then show why the Brier resolution term alone can fail to capture decision-relevant differences between sets of judgments, and finally connect this reasoning to the Blackwell ordering. The formal treatment---including proofs, the Brier decomposition, and the procedure for constructing recalibrated beliefs from data---is given in the section titled ``\nameref{sec:formal_recalibrated_resolution_decision}''.

\subsection{From Calibration to Recalibrated Elicited Beliefs}
\label{sec:recalibrated_ebs_main}

The standard approach in the psychology of probability judgment evaluates elicited beliefs using proper scoring rules, especially the Brier score, and its decomposition into reliability, resolution, and uncertainty. This decomposition is useful because it separates miscalibration from the extent to which judgments distinguish cases with different outcome frequencies. In this paper, we build directly on that tradition, but use it to motivate a richer object of analysis: the distribution of \emph{recalibrated elicited beliefs}.

To see the idea, consider our medical image classification task. Let $Y=1$ denote that an image contains a blast cell and $Y=0$ denote that it does not. A participant reports an elicited belief $p$ that $Y=1$. Suppose reports are grouped into bins indexed by $k$. In bin $k$, let $p_k$ denote the participant's reported probability, $w_k$ the fraction of observations in that bin, and $q_k$ the actual frequency of blast cells among images in that bin. Thus $p_k$ is what the participant says, while $q_k$ is what that report empirically means.

The Brier decomposition already uses both objects. The reliability term compares $p_k$ to $q_k$, measuring miscalibration. The resolution term measures how much the $q_k$'s vary around the base rate. Thus the standard decomposition implicitly replaces raw reports with their empirical meanings. We call these empirical meanings \emph{recalibrated elicited beliefs}:
\[
\tilde p_k = q_k.
\]

This transformation is important because raw reports combine two distinct features of judgment: what information the participant has extracted from the image, and how the participant maps that information onto the probability scale. Recalibration separates these. If a participant reports $.80$ on a set of images that contain blast cells only $60\%$ of the time, then for downstream decision purposes that report functions as a posterior probability of $.60$, not $.80$.

The standard Brier resolution term summarizes the recalibrated elicited beliefs by a single number:
\[
\mathrm{Resolution}
=
\sum_k w_k(q_k-\bar y)^2,
\]
where $\bar y=\sum_k w_kq_k$ is the base rate. This is simply the variance of the recalibrated elicited beliefs. That statistic is useful, but it is not the full object. The full object is the distribution of the recalibrated beliefs $\{\tilde p_k,w_k\}$.

This distinction matters in medical classification because decisions are threshold-based. A clinician may escalate, biopsy, treat, or withhold intervention depending on whether the posterior probability of malignancy exceeds a clinically relevant threshold. Such thresholds reflect the relative costs of false positives and false negatives, the risks of intervention, and the consequences of delay. Therefore, the value of a learning system depends not only on how dispersed its beliefs are, but on where its recalibrated beliefs lie relative to the relevant decision thresholds.

Our empirical analysis therefore uses recalibrated elicited beliefs for two reasons. First, they place participants on a common empirical probability scale, separating information from miscalibration. Second, their distribution is the object needed to assess what participants have learned in a decision-relevant sense. Brier resolution is one summary of this distribution; it is not a substitute for the distribution itself.

\subsection{Why Equal Resolution Need Not Mean Equal Decision Value}
\label{sec:resolution_example_main}

A simple example illustrates why the distribution of recalibrated elicited beliefs contains information that Brier resolution alone discards. The example is stated in the language of the medical image task.

Consider two perfectly calibrated learning systems for classifying whether an image contains a blast cell. Because both systems are perfectly calibrated, raw elicited beliefs and recalibrated elicited beliefs coincide.

System A produces three possible posterior probabilities:
\[
\begin{array}{c|c|c|c}
\text{Bin} & p_k & w_k & q_k \\
\hline
1 & 0 & 1/6 & 0 \\
2 & 1/2 & 2/3 & 1/2 \\
3 & 1 & 1/6 & 1
\end{array}
\]
System B produces two possible posterior probabilities. Let
\[
a=\frac12-\frac{1}{\sqrt{12}},
\qquad
b=\frac12+\frac{1}{\sqrt{12}}.
\]
Then
\[
\begin{array}{c|c|c|c}
\text{Bin} & p_k & w_k & q_k \\
\hline
1 & a & 1/2 & a \\
2 & b & 1/2 & b
\end{array}
\]

Both systems have base rate
\[
\bar y=\frac12.
\]
Both are perfectly calibrated, since $p_k=q_k$ in every bin. They also have identical Brier resolution:
\[
\mathrm{Res}_A
=
\frac16(0-1/2)^2
+
\frac23(1/2-1/2)^2
+
\frac16(1-1/2)^2
=
\frac{1}{12},
\]
and
\[
\mathrm{Res}_B
=
\frac12(a-1/2)^2
+
\frac12(b-1/2)^2
=
\frac{1}{12}.
\]
Thus, by calibration and resolution, the two systems are tied.

Now consider threshold-based clinical value. Suppose the value of a posterior belief $q$ at threshold $t$ is
\[
v_t(q)=(q-t)_+,
\]
where $(x)_+=\max\{x,0\}$. This represents the gain from having posterior probability above the intervention threshold. The ex ante value of a system is therefore
\[
V_t=\sum_k w_k(q_k-t)_+.
\]

For a low intervention threshold, $t=1/4$,
\[
V_{1/4}(A)
=
\frac16(0-1/4)_+
+
\frac23(1/2-1/4)
+
\frac16(1-1/4)
=
\frac{7}{24},
\]
whereas
\[
V_{1/4}(B)
=
\frac12(a-1/4)_+
+
\frac12(b-1/4)
=
\frac18+\frac{1}{2\sqrt{12}}.
\]
Numerically,
\[
V_{1/4}(A)\approx .2917,
\qquad
V_{1/4}(B)\approx .2693.
\]
Thus System A is better for this low-threshold decision.

For a higher intervention threshold, $t=1/2$,
\[
V_{1/2}(A)
=
\frac16(0-1/2)_+
+
\frac23(1/2-1/2)_+
+
\frac16(1-1/2)
=
\frac{1}{12},
\]
whereas
\[
V_{1/2}(B)
=
\frac12(a-1/2)_+
+
\frac12(b-1/2)
=
\frac{1}{2\sqrt{12}}.
\]
Numerically,
\[
V_{1/2}(A)\approx .0833,
\qquad
V_{1/2}(B)\approx .1443.
\]
Thus System B is better for this higher-threshold decision.

The example shows that two systems can be perfectly calibrated and have identical Brier resolution while differing in decision value. One system is better for low-threshold decisions; the other is better for high-threshold decisions. The difference lies in the placement of posterior mass relative to the decision threshold, not in overall variance around the base rate.

This motivates our use of the full distribution of recalibrated elicited beliefs. Resolution is a useful descriptive statistic, but it cannot determine which learning system is more valuable across decision contexts.

\subsection{Connection to Blackwell Comparisons}
\label{sec:blackwell_bridge_main}

The preceding example also clarifies the role of Blackwell comparisons. Blackwell's criterion asks when one information structure is more valuable than another for every possible decision problem. In a two-state setting, this is equivalent to comparing the full distributions of posterior beliefs by convex order.

The implication is straightforward. If one distribution of recalibrated elicited beliefs Blackwell-dominates another, then it supports weakly better decisions for all payoff functions. If no such dominance holds, then there is no context-free ranking: different decision problems may favor different systems.

This is not a limitation of the approach. It is the correct conclusion. In medical classification, the practical value of learning depends on the intervention threshold. When the threshold changes, the relative value of different posterior distributions can change as well. The empirical task is therefore not to force all learning systems into a single scalar ranking, but to recover the distribution of recalibrated elicited beliefs and examine how that distribution supports decisions across relevant thresholds.

The procedure implementing this comparison, and the resulting comparison of Phase 1 and Phase 3 judgments across the three conditions, are reported in the main text. The formal basis for the framework described above---including the calibration property of recalibrated beliefs, the Brier decomposition, the connection to threshold-based decisions and the Blackwell/convex order, and the procedure for constructing recalibrated beliefs from data---is given in the following section.

\section{Formal Treatment of Recalibrated Elicited Beliefs, Resolution, and Decision Value}
\label{sec:formal_recalibrated_resolution_decision}

This section provides the formal basis for our use of recalibrated elicited beliefs. The argument has five steps. First, recalibrated elicited beliefs place raw reports on an empirical probability scale. Second, the Brier decomposition shows that the standard resolution term is the variance of these recalibrated beliefs. Third, medical classification decisions are naturally threshold-based. Fourth, in a two-state setting, Blackwell comparisons correspond to convex-order comparisons of posterior distributions. Fifth, the empirical distribution of recalibrated elicited beliefs can be computed directly from observed reports and outcomes.

\subsection{Recalibrated elicited beliefs}

Let $Y\in\{0,1\}$ denote the true outcome and let $p$ denote an elicited belief that $Y=1$. Define the recalibrated elicited belief associated with report $p$ as
\[
\tilde p=\Prb(Y=1\mid p).
\]
In finite bins, if report $p_k$ is used on a set of observations in which the event occurs with frequency $q_k$, then
\[
\tilde p_k=q_k.
\]

\begin{proposition}
The recalibrated elicited belief $\tilde p$ is calibrated:
\[
\Prb(Y=1\mid \tilde p)=\tilde p.
\]
\end{proposition}

\begin{proof}
By definition, $\tilde p=\Prb(Y=1\mid p)$. Conditioning on $\tilde p$ groups together reports with the same empirical event frequency. Therefore the event frequency conditional on $\tilde p$ is $\tilde p$ itself.
\end{proof}

Raw elicited beliefs $p$ describe what participants report. Recalibrated elicited beliefs $\tilde p$ describe the empirical posterior probabilities induced by those reports. Thus recalibration separates the information encoded in a participant's judgments from distortions in the participant's use of the probability scale.

\subsection{Finite-bin Brier decomposition}

Suppose reports take finitely many values or bins $p_1,\dots,p_K$. Let $w_k$ be the fraction of observations in bin $k$, and let
\[
q_k=\Prb(Y=1\mid p=p_k)
\]
be the event frequency in that bin. Let
\[
\bar y=\sum_k w_k q_k
\]
be the overall event frequency.

The expected Brier score is
\[
\sum_k w_k\left[q_k(p_k-1)^2+(1-q_k)p_k^2\right].
\]
For each bin,
\[
q_k(p_k-1)^2+(1-q_k)p_k^2
=
p_k^2-2p_kq_k+q_k.
\]
Adding and subtracting $q_k^2$ gives
\[
p_k^2-2p_kq_k+q_k
=
(p_k-q_k)^2+q_k(1-q_k).
\]
Therefore,
\[
\mathrm{Brier}
=
\sum_k w_k(p_k-q_k)^2
+
\sum_k w_kq_k(1-q_k).
\]
Since
\[
\sum_k w_kq_k(1-q_k)
=
\bar y(1-\bar y)
-
\sum_k w_k(q_k-\bar y)^2,
\]
we obtain
\[
\mathrm{Brier}
=
\underbrace{\sum_k w_k(p_k-q_k)^2}_{\mathrm{Reliability}}
-
\underbrace{\sum_k w_k(q_k-\bar y)^2}_{\mathrm{Resolution}}
+
\underbrace{\bar y(1-\bar y)}_{\mathrm{Uncertainty}}.
\]

\begin{proposition}
The Brier resolution term is the variance of recalibrated elicited beliefs:
\[
\mathrm{Resolution}
=
\sum_k w_k(q_k-\bar y)^2
=
\Var(\tilde p).
\]
\end{proposition}

\begin{proof}
In finite bins, the recalibrated elicited belief in bin $k$ is $\tilde p_k=q_k$ and the bin weight is $w_k$. Since $\bar y=\sum_k w_kq_k$, the variance of $\tilde p$ is
\[
\sum_k w_k(\tilde p_k-\bar y)^2
=
\sum_k w_k(q_k-\bar y)^2.
\]
\end{proof}

Resolution is therefore a single second-moment summary of the distribution of recalibrated elicited beliefs. It is useful, but it is not the full distribution.

\subsection{Threshold decision value in medical classification}

Let $\theta\in\{0,1\}$ denote the true state, where $\theta=1$ corresponds to a clinically relevant positive case, such as an image containing a blast cell. Let
\[
p=\Prb(\theta=1)
\]
denote the posterior probability assigned to the positive state.

Consider two actions:
\[
a_0=\text{no intervention},
\qquad
a_1=\text{intervention}.
\]
For a threshold $t\in[0,1]$, define payoffs by
\[
u_t(a_0,\theta)=0
\qquad \text{for both } \theta\in\{0,1\},
\]
and
\[
u_t(a_1,1)=1-t,
\qquad
u_t(a_1,0)=-t.
\]
At posterior belief $p$, the expected payoff from intervention is
\[
p(1-t)+(1-p)(-t)=p-t,
\]
while the payoff from no intervention is $0$. Therefore the optimal value at posterior $p$ is
\[
v_t(p)=\max\{0,p-t\}=(p-t)_+.
\]

For a distribution $\mu$ of recalibrated elicited beliefs, the ex ante value of threshold decision problem $t$ is
\[
V_t(\mu)=\int (p-t)_+\,d\mu(p).
\]
In finite bins this becomes
\[
V_t=\sum_k w_k(q_k-t)_+.
\]

Thus, decision value depends on where the distribution of recalibrated elicited beliefs lies relative to the threshold $t$. Different clinical thresholds can favor different posterior distributions, even when those distributions have the same mean and variance.

\subsection{Blackwell comparisons and convex order}

Let $\mu$ and $\nu$ be two distributions of posterior beliefs on $[0,1]$ with the same mean. We say that $\nu$ dominates $\mu$ in convex order, written
\[
\mu\preceq_{cx}\nu,
\]
if
\[
\int \phi(p)\,d\mu(p)
\le
\int \phi(p)\,d\nu(p)
\]
for every convex function $\phi:[0,1]\to\R$.

In a two-state environment, this convex-order comparison is equivalent to the Blackwell comparison of the corresponding information structures: the more informative structure induces the posterior distribution that is larger in convex order.

For posterior distributions with the same mean, convex order is equivalently characterized by the threshold, or hinge, functions
\[
(p-t)_+.
\]
That is,
\[
\mu\preceq_{cx}\nu
\]
if and only if
\[
\int (p-t)_+\,d\mu(p)
\le
\int (p-t)_+\,d\nu(p)
\qquad
\text{for all } t\in[0,1].
\]
In finite support form, this condition is
\[
\sum_i a_i(p_i-t)_+
\le
\sum_j b_j(q_j-t)_+
\qquad
\text{for all } t\in[0,1].
\]

This characterization is especially useful here because the hinge function is exactly the threshold decision value derived above. If one distribution of recalibrated elicited beliefs dominates another in convex order, it yields weakly higher value for every threshold decision problem. If the hinge inequalities cross, then neither system is uniformly more informative: one may be better for some treatment thresholds, while the other may be better for others.

\subsection{Same resolution, opposite threshold rankings}

The following example shows why Brier resolution alone cannot determine decision value. Let
\[
\mu_A
=
\frac16\delta_0+\frac23\delta_{1/2}+\frac16\delta_1
\]
and
\[
\mu_B
=
\frac12\delta_{1/2-1/\sqrt{12}}
+
\frac12\delta_{1/2+1/\sqrt{12}}.
\]
Here $\delta_x$ denotes a point mass at $x$.

Both distributions have mean $1/2$. Distribution $\mu_A$ has variance
\[
\frac16(0-1/2)^2
+
\frac23(1/2-1/2)^2
+
\frac16(1-1/2)^2
=
\frac{1}{12}.
\]
Distribution $\mu_B$ has variance
\[
\frac12\left(1/2-1/\sqrt{12}-1/2\right)^2
+
\frac12\left(1/2+1/\sqrt{12}-1/2\right)^2
=
\frac{1}{12}.
\]
Thus the two distributions have the same Brier resolution.

However, their threshold values differ. For $t=1/4$,
\[
\int(p-1/4)_+\,d\mu_A(p)
=
\frac16(0-1/4)_+
+
\frac23(1/2-1/4)
+
\frac16(1-1/4)
=
\frac{7}{24},
\]
while
\[
\int(p-1/4)_+\,d\mu_B(p)
=
\frac12(1/2-1/\sqrt{12}-1/4)_+
+
\frac12(1/2+1/\sqrt{12}-1/4)
=
\frac18+\frac{1}{2\sqrt{12}}.
\]
Since
\[
\frac{7}{24}
>
\frac18+\frac{1}{2\sqrt{12}},
\]
$\mu_A$ yields higher value at threshold $t=1/4$.

For $t=1/2$,
\[
\int(p-1/2)_+\,d\mu_A(p)
=
\frac16(1-1/2)
=
\frac{1}{12},
\]
while
\[
\int(p-1/2)_+\,d\mu_B(p)
=
\frac12(1/2+1/\sqrt{12}-1/2)
=
\frac{1}{2\sqrt{12}}.
\]
Since
\[
\frac{1}{12}
<
\frac{1}{2\sqrt{12}},
\]
$\mu_B$ yields higher value at threshold $t=1/2$.

Thus two posterior distributions can have the same mean and the same variance, and hence the same Brier resolution, while reversing their rankings across threshold decision problems. This is why resolution does not determine decision value. The relevant object is the full distribution of recalibrated elicited beliefs.

\subsection{Empirical construction of recalibrated elicited beliefs}

We now describe the in-principle computational procedure for constructing the empirical distribution of recalibrated elicited beliefs.

Suppose the data consist of observations indexed by $i=1,\dots,N$. For each observation, we observe:
\[
(p_i,Y_i),
\]
where $p_i\in[0,1]$ is the raw elicited belief and $Y_i\in\{0,1\}$ is the true outcome.

Choose a partition of the unit interval into bins
\[
B_1,\dots,B_K.
\]
For each bin $B_k$, define
\[
n_k=\#\{i:p_i\in B_k\},
\]
and
\[
w_k=\frac{n_k}{N}.
\]
The empirical event frequency in bin $k$ is
\[
q_k=
\frac{1}{n_k}
\sum_{i:p_i\in B_k}Y_i,
\]
provided $n_k>0$. The recalibrated elicited belief associated with bin $k$ is
\[
\tilde p_k=q_k.
\]
The empirical distribution of recalibrated elicited beliefs is therefore
\[
\widehat{\mu}_{\tilde p}
=
\{(\tilde p_k,w_k):k=1,\dots,K,\ n_k>0\}.
\]

This construction can be applied separately by experimental condition, phase, participant group, or individual participant, depending on sample size. For example, if $g$ indexes an experimental condition, then
\[
q_{kg}
=
\frac{1}{n_{kg}}
\sum_{i:p_i\in B_k,\ i\in g}Y_i,
\qquad
w_{kg}
=
\frac{n_{kg}}{N_g}.
\]
The condition-specific distribution of recalibrated elicited beliefs is
\[
\widehat{\mu}_{\tilde p,g}
=
\{(q_{kg},w_{kg}):k=1,\dots,K,\ n_{kg}>0\}.
\]

Once $\widehat{\mu}_{\tilde p,g}$ is constructed, the Brier resolution for condition $g$ is
\[
\widehat{\mathrm{Resolution}}_g
=
\sum_k w_{kg}(q_{kg}-\bar y_g)^2,
\]
where
\[
\bar y_g=\sum_k w_{kg}q_{kg}.
\]
The threshold value curve for condition $g$ is
\[
\widehat V_g(t)
=
\sum_k w_{kg}(q_{kg}-t)_+.
\]

Plotting $\widehat V_g(t)$ over thresholds $t\in[0,1]$ shows which condition yields greater decision value at each threshold. If
\[
\widehat V_g(t)\ge \widehat V_h(t)
\qquad
\text{for all } t\in[0,1],
\]
then condition $g$ dominates condition $h$ in the empirical hinge-function sense. If the curves cross, then neither condition is uniformly more decision-relevant; different thresholds favor different conditions.

In discrete empirical applications, it is enough to evaluate the hinge inequalities on the finite set of observed recalibrated belief values and the endpoints:
\[
t\in
\{0,1\}
\cup
\{q_{kg}:n_{kg}>0\}
\cup
\{q_{kh}:n_{kh}>0\}.
\]
Between these points, the threshold value functions are linear.

\subsection{Algorithmic summary}

The empirical procedure can be summarized as follows.

\begin{enumerate}
\item \textbf{Choose the comparison level.} Decide whether recalibrated elicited beliefs will be constructed by condition, phase, participant group, or individual participant.

\item \textbf{Bin raw elicited beliefs.} Partition $[0,1]$ into bins $B_1,\dots,B_K$. The same binning rule should be used across compared groups.

\item \textbf{Compute empirical meanings.} For each group $g$ and bin $k$, compute the empirical event frequency
\[
q_{kg}
=
\frac{1}{n_{kg}}
\sum_{i:p_i\in B_k,\ i\in g}Y_i.
\]

\item \textbf{Construct recalibrated EB distributions.} Form
\[
\widehat{\mu}_{\tilde p,g}
=
\{(q_{kg},w_{kg}):n_{kg}>0\}.
\]

\item \textbf{Compute Brier components.} Reliability, resolution, and uncertainty can be computed as
\[
\widehat{\mathrm{Reliability}}_g
=
\sum_k w_{kg}(p_{kg}-q_{kg})^2,
\]
\[
\widehat{\mathrm{Resolution}}_g
=
\sum_k w_{kg}(q_{kg}-\bar y_g)^2,
\]
and
\[
\widehat{\mathrm{Uncertainty}}_g
=
\bar y_g(1-\bar y_g),
\]
where $p_{kg}$ is the representative reported probability for bin $k$ in group $g$.

\item \textbf{Compute threshold value curves.} For each group $g$, compute
\[
\widehat V_g(t)
=
\sum_k w_{kg}(q_{kg}-t)_+
\]
over a grid of thresholds $t\in[0,1]$.

\item \textbf{Compare groups.} If one group has weakly larger $\widehat V_g(t)$ at every threshold, it is empirically more valuable across all threshold decision problems. If the curves cross, the comparison is threshold-dependent.

\item \textbf{Report interpretation.} Resolution describes the variance of recalibrated elicited beliefs. The threshold value curve describes how the full distribution of recalibrated elicited beliefs supports medical decisions across thresholds.
\end{enumerate}

This procedure makes explicit the distinction between standard Brier-based summaries and the decision-relevant distributional object used in the paper.

\bibliographystyle{unsrtnat}
\bibliography{bibliography_bibtex}

\begin{thebibliography}{40}
\providecommand{\natexlab}[1]{#1}
\providecommand{\url}[1]{\texttt{#1}}
\expandafter\ifx\csname urlstyle\endcsname\relax
  \providecommand{\doi}[1]{doi: #1}\else
  \providecommand{\doi}{doi: \begingroup \urlstyle{rm}\Url}\fi

\bibitem[Steyvers and
  Kumar(2024)]{Steyvers.Kumar.2024_ThreeChallengesAIAssisted}
Mark Steyvers and Aakriti Kumar.
\newblock Three {{Challenges}} for {{AI-Assisted Decision-Making}}.
\newblock \emph{Perspectives on Psychological Science}, 19\penalty0
  (5):\penalty0 722--734, 2024.
\newblock \doi{10.1177/17456916231181102}.
\newblock URL \url{https://journals.sagepub.com/doi/10.1177/17456916231181102}.

\bibitem[Ma et~al.(2024)Ma, Wang, Lei, Shi, Yin, and
  Ma]{Ma.et.al.2024_AreYouReally}
Shuai Ma, Xinru Wang, Ying Lei, Chuhan Shi, Ming Yin, and Xiaojuan Ma.
\newblock “{{Are You Really Sure}}?” {{Understanding}} the {{Effects}} of
  {{Human Self-Confidence Calibration}} in {{AI-Assisted Decision Making}}.
\newblock In \emph{Proceedings of the {{CHI Conference}} on {{Human Factors}}
  in {{Computing Systems}}}, pages 1--20, Honolulu HI USA, 2024. ACM.
\newblock \doi{10.1145/3613904.3642671}.
\newblock URL \url{https://dl.acm.org/doi/10.1145/3613904.3642671}.

\bibitem[Caplin et~al.(2025)Caplin, Deming, Li, Martin, Marx, Weidmann, and
  Ye]{Caplin.et.al.2025_ABCsWhoBenefits}
Andrew Caplin, David Deming, Shangwen Li, Daniel Martin, Philip Marx, Ben
  Weidmann, and Kadachi~Jiada Ye.
\newblock The {{ABCs}} of {{Who Benefits}} from {{Working}} with {{AI}}:
  {{Ability}}, {{Beliefs}}, and {{Calibration}}.
\newblock \emph{Management Science}, page mnsc.2024.08994, 2025.
\newblock \doi{10.1287/mnsc.2024.08994}.
\newblock URL \url{https://pubsonline.informs.org/doi/10.1287/mnsc.2024.08994}.

\bibitem[Epping et~al.(2026{\natexlab{a}})Epping, Caplin, Duhaime, Holmes,
  Martin, and Trueblood]{Epping.et.al.2026_ImprovingCrowdsourcingAI}
Gunnar~P. Epping, Andrew Caplin, Erik Duhaime, William~R. Holmes, Daniel
  Martin, and Jennifer~S. Trueblood.
\newblock Improving crowdsourcing for {{AI}} through cognitive-inspired data
  engineering.
\newblock \emph{Behavior Research Methods}, 58\penalty0 (10):\penalty0 283,
  2026{\natexlab{a}}.
\newblock ISSN 1554-3528.
\newblock \doi{10.3758/s13428-026-03160-4}.
\newblock URL \url{https://link.springer.com/10.3758/s13428-026-03160-4}.

\bibitem[Epping et~al.(2026{\natexlab{b}})Epping, Caplin, Duhaime, Holmes,
  Martin, and Trueblood]{Epping.et.al.2025_HarnessingHumanUncertainty}
Gunnar~P Epping, Andrew Caplin, Erik Duhaime, William~R Holmes, Daniel Martin,
  and Jennifer~S Trueblood.
\newblock Harnessing human uncertainty to train more accurate and aligned ai
  systems.
\newblock \emph{Decision Analysis}, 2026{\natexlab{b}}.
\newblock \doi{https://doi.org/10.1287/deca.2025.0395}.

\bibitem[Long and Magerko(2020)]{Long.Magerko.2020_WhatAILiteracy}
Duri Long and Brian Magerko.
\newblock What is {AI} literacy? {Competencies} and design considerations.
\newblock In \emph{Proceedings of the 2020 {CHI} Conference on Human Factors in
  Computing Systems}, pages 1--16, Honolulu, HI, USA, 2020. ACM.
\newblock \doi{10.1145/3313831.3376727}.

\bibitem[Lichtenstein et~al.(1982)Lichtenstein, Fischhoff, and
  Phillips]{Lichtenstein.et.al.1982_CalibrationProbabilitiesState}
Sarah Lichtenstein, Baruch Fischhoff, and Lawrence~D. Phillips.
\newblock Calibration of probabilities: {{The}} state of the art to 1980.
\newblock In Daniel Kahneman, Paul Slovic, and Amos Tversky, editors,
  \emph{Judgment under {{Uncertainty}}}, pages 306--334. Cambridge University
  Press, 1 edition, 1982.
\newblock \doi{10.1017/CBO9780511809477.023}.
\newblock URL
  \url{https://www.cambridge.org/core/product/identifier/CBO9780511809477A036/type/book_part}.

\bibitem[Griffin and
  Brenner(2004)]{Griffin.Brenner.2004_PerspectivesProbabilityJudgment}
Dale Griffin and Lyle Brenner.
\newblock Perspectives on {{Probability Judgment Calibration}}.
\newblock In Derek~J. Koehler and Nigel Harvey, editors, \emph{Blackwell
  {{Handbook}} of {{Judgment}} and {{Decision Making}}}, pages 177--199. Wiley,
  1 edition, 2004.
\newblock \doi{10.1002/9780470752937.ch9}.
\newblock URL
  \url{https://onlinelibrary.wiley.com/doi/10.1002/9780470752937.ch9}.

\bibitem[Guo et~al.(2017)Guo, Pleiss, Sun, and
  Weinberger]{Guo.et.al.2017_CalibrationModernNeural}
Chuan Guo, Geoff Pleiss, Yu~Sun, and Kilian~Q. Weinberger.
\newblock On calibration of modern neural networks.
\newblock In Doina Precup and Yee~Whye Teh, editors, \emph{Proceedings of the
  34th International Conference on Machine Learning}, volume~70 of
  \emph{Proceedings of Machine Learning Research}, pages 1321--1330. PMLR,
  2017.
\newblock URL \url{https://proceedings.mlr.press/v70/guo17a.html}.

\bibitem[Lichtenstein and
  Fischhoff(1980)]{Lichtenstein.Fischhoff.1980_TrainingCalibration}
Sarah Lichtenstein and Baruch Fischhoff.
\newblock Training for calibration.
\newblock \emph{Organizational Behavior and Human Performance}, 26\penalty0
  (2):\penalty0 149--171, 1980.
\newblock \doi{10.1016/0030-5073(80)90052-5}.
\newblock URL
  \url{https://linkinghub.elsevier.com/retrieve/pii/0030507380900525}.

\bibitem[Stone and Opel(2000)]{Stone.Opel.2000_TrainingImproveCalibration}
Eric~R. Stone and Ryan~B. Opel.
\newblock Training to {{Improve Calibration}} and {{Discrimination}}: {{The
  Effects}} of {{Performance}} and {{Environmental Feedback}}.
\newblock \emph{Organizational Behavior and Human Decision Processes},
  83\penalty0 (2):\penalty0 282--309, 2000.
\newblock \doi{10.1006/obhd.2000.2910}.
\newblock URL
  \url{https://linkinghub.elsevier.com/retrieve/pii/S0749597800929108}.

\bibitem[Sharp et~al.(1988)Sharp, Cutler, and
  Penrod]{Sharp.et.al.1988_PerformanceFeedbackImproves}
Glen~L Sharp, Brian~L Cutler, and Steven~D Penrod.
\newblock Performance feedback improves the resolution of confidence judgments.
\newblock \emph{Organizational Behavior and Human Decision Processes},
  42\penalty0 (3):\penalty0 271--283, 1988.
\newblock \doi{10.1016/0749-5978(88)90001-5}.
\newblock URL
  \url{https://linkinghub.elsevier.com/retrieve/pii/0749597888900015}.

\bibitem[Kelly and Mandel(2024)]{Kelly.Mandel.2024_EffectCalibrationTraining}
Megan~O. Kelly and David~R. Mandel.
\newblock The effect of calibration training on the calibration of intelligence
  analysts' judgments.
\newblock \emph{Applied Cognitive Psychology}, 38\penalty0 (5):\penalty0 e4236,
  2024.
\newblock \doi{10.1002/acp.4236}.
\newblock URL \url{https://onlinelibrary.wiley.com/doi/10.1002/acp.4236}.

\bibitem[Stone et~al.(2023)Stone, Luu, Costello, and
  Somerville]{Stone.et.al.2023_AutomatedCalibrationTraining}
Eric~R. Stone, Jason Luu, Cory~K. Costello, and Annie~H. Somerville.
\newblock Automated calibration training for forecasters.
\newblock \emph{Journal of Behavioral Decision Making}, 36\penalty0
  (4):\penalty0 e2334, 2023.
\newblock \doi{10.1002/bdm.2334}.
\newblock URL \url{https://onlinelibrary.wiley.com/doi/10.1002/bdm.2334}.

\bibitem[Moore et~al.(2017)Moore, Swift, Minster, Mellers, Ungar, Tetlock,
  Yang, and Tenney]{Moore.et.al.2017_ConfidenceCalibrationMultiyear}
Don~A. Moore, Samuel~A. Swift, Angela Minster, Barbara Mellers, Lyle Ungar,
  Philip Tetlock, Heather H.~J. Yang, and Elizabeth~R. Tenney.
\newblock Confidence {{Calibration}} in a {{Multiyear Geopolitical Forecasting
  Competition}}.
\newblock \emph{Management Science}, 63\penalty0 (11):\penalty0 3552--3565,
  2017.
\newblock \doi{10.1287/mnsc.2016.2525}.
\newblock URL \url{https://pubsonline.informs.org/doi/10.1287/mnsc.2016.2525}.

\bibitem[Gruetzemacher et~al.(2024)Gruetzemacher, Lee, and
  Paradice]{Gruetzemacher.et.al.2024_CalibrationTrainingImproving}
Ross Gruetzemacher, Kang~Bok Lee, and David Paradice.
\newblock Calibration training for improving probabilistic judgments using an
  interactive app.
\newblock \emph{FUTURES \& FORESIGHT SCIENCE}, 6\penalty0 (2):\penalty0 e177,
  2024.
\newblock \doi{10.1002/ffo2.177}.
\newblock URL \url{https://onlinelibrary.wiley.com/doi/10.1002/ffo2.177}.

\bibitem[Kumar et~al.(2019)Kumar, Liang, and
  Ma]{Kumar.et.al.2019_VerifiedUncertaintyCalibration}
Ananya Kumar, Percy~S Liang, and Tengyu Ma.
\newblock Verified uncertainty calibration.
\newblock In H.~Wallach, H.~Larochelle, A.~Beygelzimer, F.~{d'Alch\'e-Buc},
  E.~Fox, and R.~Garnett, editors, \emph{Advances in Neural Information
  Processing Systems}, volume~32. Curran Associates, Inc., 2019.
\newblock URL
  \url{https://proceedings.neurips.cc/paper_files/paper/2019/file/f8c0c968632845cd133308b1a494967f-Paper.pdf}.

\bibitem[Baranski and
  Petrusic(1994)]{Baranski.Petrusic.1994_CalibrationResolutionConfidence}
Joseph~V. Baranski and William~M. Petrusic.
\newblock The calibration and resolution of confidence in perceptual judgments.
\newblock \emph{Perception \& Psychophysics}, 55\penalty0 (4):\penalty0
  412--428, 1994.
\newblock \doi{10.3758/BF03205299}.
\newblock URL \url{http://link.springer.com/10.3758/BF03205299}.

\bibitem[Martin and
  Mandel(2025)]{Martin.Mandel.2025_CalibrationFeedbackPractical}
Matthew Martin and David~R. Mandel.
\newblock Calibration {{Feedback With}} the {{Practical Scoring Rule Does Not
  Improve Calibration}} of {{Confidence}}.
\newblock \emph{FUTURES \& FORESIGHT SCIENCE}, 7\penalty0 (1):\penalty0 e199,
  2025.
\newblock \doi{10.1002/ffo2.199}.
\newblock URL \url{https://onlinelibrary.wiley.com/doi/10.1002/ffo2.199}.

\bibitem[Vodrahalli et~al.(2022)Vodrahalli, Gerstenberg, and
  Zou]{Vodrahalli.et.al.2022_UncalibratedModelsCan}
Kailas Vodrahalli, Tobias Gerstenberg, and James~Y Zou.
\newblock Uncalibrated models can improve human-{{AI}} collaboration.
\newblock In S.~Koyejo, S.~Mohamed, A.~Agarwal, D.~Belgrave, K.~Cho, and A.~Oh,
  editors, \emph{Advances in Neural Information Processing Systems}, volume~35,
  pages 4004--4016. Curran Associates, Inc., 2022.
\newblock URL
  \url{https://proceedings.neurips.cc/paper_files/paper/2022/file/1968ea7d985aa377e3a610b05fc79be0-Paper-Conference.pdf}.

\bibitem[Corvelo~Benz and Gomez~Rodriguez(2023)]{benz2023humanaligned}
Nina~L. Corvelo~Benz and Manuel Gomez~Rodriguez.
\newblock Human-aligned calibration for {{AI-assisted}} decision making.
\newblock In \emph{Thirty-Seventh Conference on Neural Information Processing
  Systems}, 2023.
\newblock URL \url{https://openreview.net/forum?id=GfITbjrIOd}.

\bibitem[Noti et~al.(2025)Noti, Donahue, Kleinberg, and
  Oren]{Noti.et.al.2025_AIAssistedDecisionHumanLearning}
Gali Noti, Kate Donahue, Jon Kleinberg, and Sigal Oren.
\newblock {AI}-assisted decision making with human learning.
\newblock In \emph{Proceedings of the 26th {ACM} Conference on Economics and
  Computation ({EC} '25)}, Stanford, CA, USA, 2025. ACM.
\newblock \doi{10.1145/3736252.3742492}.

\bibitem[Li et~al.(2025)Li, Yang, Liao, Zhang, and
  Lee]{Li.et.al.2025_ConfidenceAlignsUnderstanding}
Jingshu Li, Yitian Yang, Q.~Vera Liao, Junti Zhang, and Yi-Chieh Lee.
\newblock As {{Confidence Aligns}}: {{Understanding}} the {{Effect}} of {{AI
  Confidence}} on {{Human Self-confidence}} in {{Human-AI Decision Making}}.
\newblock In \emph{Proceedings of the 2025 {{CHI Conference}} on {{Human
  Factors}} in {{Computing Systems}}}, pages 1--16, Yokohama Japan, 2025. ACM.
\newblock \doi{10.1145/3706598.3713336}.
\newblock URL \url{https://dl.acm.org/doi/10.1145/3706598.3713336}.

\bibitem[Hadfield-Menell et~al.(2016)Hadfield-Menell, Russell, Abbeel, and
  Dragan]{HadfieldMenell.et.al.2016_CooperativeInverse}
Dylan Hadfield-Menell, Stuart~J. Russell, Pieter Abbeel, and Anca~D. Dragan.
\newblock Cooperative inverse reinforcement learning.
\newblock In \emph{Advances in Neural Information Processing Systems},
  volume~29, pages 3909--3917. Curran Associates, Inc., 2016.

\bibitem[Murphy(1973)]{Murphy.1973_NewVectorPartition}
Allan~H. Murphy.
\newblock A {{New Vector Partition}} of the {{Probability Score}}.
\newblock \emph{Journal of Applied Meteorology}, 12\penalty0 (4):\penalty0
  595--600, 1973.
\newblock \doi{10.1175/1520-0450(1973)012<0595:ANVPOT>2.0.CO;2}.
\newblock URL
  \url{http://journals.ametsoc.org/doi/10.1175/1520-0450(1973)012<0595:ANVPOT>2.0.CO;2}.

\bibitem[Turner et~al.(2014)Turner, Steyvers, Merkle, Budescu, and
  Wallsten]{Turner.et.al.2014_ForecastAggregationRecalibration}
Brandon~M. Turner, Mark Steyvers, Edgar~C. Merkle, David~V. Budescu, and
  Thomas~S. Wallsten.
\newblock Forecast aggregation via recalibration.
\newblock \emph{Machine Learning}, 95\penalty0 (3):\penalty0 261--289, 2014.
\newblock \doi{10.1007/s10994-013-5401-4}.
\newblock URL \url{http://link.springer.com/10.1007/s10994-013-5401-4}.

\bibitem[Blackwell(1951)]{Blackwell.1951_ComparisonExperiments}
David Blackwell.
\newblock Comparison of {{Experiments}}.
\newblock In Jerzy Neyman, editor, \emph{Proceedings of the {{Second Berkeley
  Symposium}} on {{Mathematical Statistics}} and {{Probability}}}, pages
  93--102. University of California Press, 1951.
\newblock \doi{10.1525/9780520411586-009}.
\newblock URL
  \url{https://www.degruyter.com/document/doi/10.1525/9780520411586-009/html}.

\bibitem[Keren(1991)]{Keren.1991_CalibrationProbabilityJudgements}
Gideon Keren.
\newblock Calibration and probability judgements: {{Conceptual}} and
  methodological issues.
\newblock \emph{Acta Psychologica}, 77\penalty0 (3):\penalty0 217--273, 1991.
\newblock \doi{10.1016/0001-6918(91)90036-Y}.
\newblock URL
  \url{https://linkinghub.elsevier.com/retrieve/pii/000169189190036Y}.

\bibitem[Corvelo~Benz and
  Gomez~Rodriguez(2025)]{CorveloBenz.GomezRodriguez.2025_HumanalignmentInfluencesUtility}
Nina~L. Corvelo~Benz and Manuel Gomez~Rodriguez.
\newblock Human-alignment influences the utility of {{AI-assisted}} decision
  making.
\newblock \emph{Scientific Reports}, 15\penalty0 (1):\penalty0 29154, 2025.
\newblock \doi{10.1038/s41598-025-12205-1}.
\newblock URL \url{https://www.nature.com/articles/s41598-025-12205-1}.

\bibitem[Trueblood et~al.(2018)Trueblood, Holmes, Seegmiller, Douds, Compton,
  Szentirmai, Woodruff, Huang, Stratton, and
  Eichbaum]{Trueblood.et.al.2018_ImpactSpeedBias}
Jennifer~S. Trueblood, William~R. Holmes, Adam~C. Seegmiller, Jonathan Douds,
  Margaret Compton, Eszter Szentirmai, Megan Woodruff, Wenrui Huang, Charles
  Stratton, and Quentin Eichbaum.
\newblock The impact of speed and bias on the cognitive processes of experts
  and novices in medical image decision-making.
\newblock \emph{Cognitive Research: Principles and Implications}, 3\penalty0
  (1):\penalty0 28, 2018.
\newblock \doi{10.1186/s41235-018-0119-2}.
\newblock URL \url{https://doi.org/10.1186/s41235-018-0119-2}.

\bibitem[Hasan et~al.(2024)Hasan, Eichbaum, Seegmiller, Stratton, and
  Trueblood]{hasan2024harnessing}
Eeshan Hasan, Quentin Eichbaum, Adam~C Seegmiller, Charles Stratton, and
  Jennifer~S Trueblood.
\newblock Harnessing the wisdom of the confident crowd in medical image
  decision-making.
\newblock \emph{Decision}, 11\penalty0 (1):\penalty0 127, 2024.

\bibitem[De~Leeuw et~al.(2023)De~Leeuw, Gilbert, and
  Luchterhandt]{de2023jspsych}
Joshua~R De~Leeuw, Rebecca~A Gilbert, and Bj{\"o}rn Luchterhandt.
\newblock jspsych: Enabling an open-source collaborative ecosystem of
  behavioral experiments.
\newblock \emph{Journal of Open Source Software}, 8\penalty0 (85):\penalty0
  5351, 2023.
\newblock \doi{10.21105/joss.05351}.

\bibitem[{Open Science Tools}()]{Pavlovia}
{Open Science Tools}.
\newblock Pavlovia.
\newblock \url{https://pavlovia.org}.
\newblock URL \url{https://pavlovia.org}.

\bibitem[Tsutsui et~al.(2023)Tsutsui, Pang, and
  Wen]{Tsutsui.et.al.2023_WBCAttWhiteBlood}
Satoshi Tsutsui, Winnie Pang, and Bihan Wen.
\newblock {{WBCAtt}}: {{A White Blood Cell Dataset Annotated}} with {{Detailed
  Morphological Attributes}}, 2023.
\newblock URL \url{http://arxiv.org/abs/2306.13531}.

\bibitem[{Stan Development
  Team}(2025)]{StanDevelopmentTeam.2025_StanReferenceManual}
{Stan Development Team}.
\newblock Stan {{Reference Manual}}, 2025.
\newblock URL \url{https://mc-stan.org}.

\bibitem[Gabry et~al.(2025)Gabry, Češnovar, Johnson, and
  Bronder]{Gabry.et.al.2025_CmdstanrInterfaceCmdStan}
Jonah Gabry, Rok Češnovar, Andrew Johnson, and Steve Bronder.
\newblock Cmdstanr: {{R Interface}} to '{{CmdStan}}', 2025.
\newblock URL \url{https://mc-stan.org/cmdstanr/}.

\bibitem[Bürkner(2018)]{Burkner.2018_AdvancedBayesianMultilevel}
Paul-Christian Bürkner.
\newblock Advanced {{Bayesian Multilevel Modeling}} with the {{R Package}}
  brms.
\newblock \emph{The R Journal}, 10\penalty0 (1):\penalty0 395, 2018.
\newblock \doi{10.32614/RJ-2018-017}.
\newblock URL
  \url{https://journal.r-project.org/archive/2018/RJ-2018-017/index.html}.

\bibitem[Arel-Bundock(2026)]{Arel-Bundock.2026_ModelMeaningHow}
Vincent Arel-Bundock.
\newblock \emph{Model to {{Meaning}}: {{How}} to {{Interpret Statistical
  Models}} with {{R}} and {{Python}}}.
\newblock CHAPMAN \& HALL CRC, S.l., 2026.
\newblock \doi{10.1201/9781003560333}.

\bibitem[Tversky and Fox(1995)]{Tversky.Fox.1995_WeighingRiskUncertainty}
Amos Tversky and Craig~R. Fox.
\newblock Weighing risk and uncertainty.
\newblock \emph{Psychological Review}, 102\penalty0 (2):\penalty0 269--283,
  1995.
\newblock \doi{10.1037/0033-295X.102.2.269}.
\newblock URL \url{https://doi.apa.org/doi/10.1037/0033-295X.102.2.269}.

\bibitem[Gonzalez and Wu(1999)]{Gonzalez.Wu.1999_ShapeProbabilityWeighting}
Richard Gonzalez and George Wu.
\newblock On the {{Shape}} of the {{Probability Weighting Function}}.
\newblock \emph{Cognitive Psychology}, 38\penalty0 (1):\penalty0 129--166,
  1999.
\newblock \doi{10.1006/cogp.1998.0710}.
\newblock URL
  \url{https://linkinghub.elsevier.com/retrieve/pii/S0010028598907101}.

\end{thebibliography}


\begin{thebibliography}{4}
\providecommand{\natexlab}[1]{#1}
\providecommand{\url}[1]{\texttt{#1}}
\expandafter\ifx\csname urlstyle\endcsname\relax
  \providecommand{\doi}[1]{doi: #1}\else
  \providecommand{\doi}{doi: \begingroup \urlstyle{rm}\Url}\fi

\bibitem[B{\"u}rkner(2017)]{Burkner.2017_BrmsPackageBayesian}
Paul-Christian B{\"u}rkner.
\newblock brms: An {R} package for {Bayesian} multilevel models using {Stan}.
\newblock \emph{Journal of Statistical Software}, 80\penalty0 (1):\penalty0
  1--28, 2017.
\newblock \doi{10.18637/jss.v080.i01}.
\newblock URL \url{http://www.jstatsoft.org/v80/i01/}.

\bibitem[Gabry et~al.(2025)Gabry, Češnovar, Johnson, and
  Bronder]{Gabry.et.al.2025_CmdstanrInterfaceCmdStan}
Jonah Gabry, Rok Češnovar, Andrew Johnson, and Steve Bronder.
\newblock Cmdstanr: {{R Interface}} to '{{CmdStan}}', 2025.
\newblock URL \url{https://mc-stan.org/cmdstanr/}.

\bibitem[Epping et~al.(2026)Epping, Caplin, Duhaime, Holmes, Martin, and
  Trueblood]{Epping.et.al.2026_ImprovingCrowdsourcingAI}
Gunnar~P. Epping, Andrew Caplin, Erik Duhaime, William~R. Holmes, Daniel
  Martin, and Jennifer~S. Trueblood.
\newblock Improving crowdsourcing for {{AI}} through cognitive-inspired data
  engineering.
\newblock \emph{Behavior Research Methods}, 58\penalty0 (10):\penalty0 283,
  2026.
\newblock ISSN 1554-3528.
\newblock \doi{10.3758/s13428-026-03160-4}.
\newblock URL \url{https://link.springer.com/10.3758/s13428-026-03160-4}.

\bibitem[Tsutsui et~al.(2023)Tsutsui, Pang, and
  Wen]{Tsutsui.et.al.2023_WBCAttWhiteBlood}
Satoshi Tsutsui, Winnie Pang, and Bihan Wen.
\newblock {{WBCAtt}}: {{A White Blood Cell Dataset Annotated}} with {{Detailed
  Morphological Attributes}}, 2023.
\newblock URL \url{http://arxiv.org/abs/2306.13531}.

\end{thebibliography}
\let\addcontentsline\articleAddContentsLine

\end{document}